\documentclass[%
 reprint,
 amsmath,
 amssymb,
 aps,
 prb,
 superscriptaddress,
 nofootinbib,
]{revtex4-2}

\usepackage{graphicx}
\usepackage{dcolumn}
\usepackage{bm}
\newcommand{\ket}[1]{|{#1}\rangle}

\newcommand{\braket}[1]{\langle{#1}\rangle}
\usepackage[caption=false]{subfig}

\usepackage{mathtools}
\usepackage[T1]{fontenc}
\usepackage[utf8]{inputenc}
\usepackage{textcomp}
\usepackage[english]{babel}
\usepackage{lmodern}
\usepackage[babel]{microtype}

\DeclareUnicodeCharacter{2013}{--} 

\usepackage{xcolor}
\usepackage[colorlinks,
			linkcolor=black,
			citecolor=black,
			filecolor=black,
			urlcolor=blue!50!black,
			pdfusetitle]{hyperref}
\hypersetup{
	pdfauthor={Axel Fünfhaus, Marius Möller, Thilo Kopp, Roser Valent\'i},
	pdfsubject={Solid-State Physics},
	pdfkeywords={topology, magnetic field, interacting phases, Hofstadter model}
 }
\let\vec\boldsymbol

\def\ci{\mathrm{i}}
\DeclareMathOperator{\e}{e}

\begin{document}


\title{Band representations in Strongly Correlated Settings: The Kitaev Honeycomb Model}

\author{Axel Fünfhaus}
\email{fuenfhaus@itp.uni-frankfurt.de}
\affiliation{
 Institute of Theoretical Physics, Goethe University Frankfurt, Max-von-Laue-Straße 1, 60438 Frankfurt am Main, Germany
}
\author{Mikel Garc\'ia-D\'iez}
\affiliation{
 Donostia International Physics Center, 20018 Donostia-San Sebasti\'an, Spain
}
\affiliation{
 Physics Department, University of the Basque Country (UPV/EHU), Bilbao, Spain
}
\author{M. G. Vergniory}
\affiliation{
 Donostia International Physics Center, 20018 Donostia-San Sebasti\'an, Spain
}
\affiliation{
Département de Physique et Institut Quantique,
Université de Sherbrooke, Sherbrooke, J1K 2R1 Québec, Canada.
}

\author{Thilo Kopp}
\affiliation{
 Center for Electronic Correlations and Magnetism, Institute of Physics, University of Augsburg, 86135 Augsburg, Germany
}
\author{Stephen M. Winter}
\affiliation{
Department of Physics and Center for Functional Materials, Wake Forest University, Winston-Salem, North Carolina 27109, USA
}
\author{Roser Valent\'i}
\affiliation{
 Institute of Theoretical Physics, Goethe University Frankfurt, Max-von-Laue-Straße 1, 60438 Frankfurt am Main, Germany
}

\date{\today}

\begin{abstract}
In the study of quantum spin liquids, the Kitaev model plays a pivotal role due to the fact that its ground state is exactly known as well as the fact that it may be realized in strongly frustrated materials such as ${\alpha}$-RuCl${}_3$. While topological insulators and superconductors can be investigated by means of topological band theory --- in particular the topological quantum chemistry (TQC) formalism --- the Kitaev model evades such a treatment, as it is not possible to set up a proper single-particle Green’s function for it. We instead associate spin operators with ``orbitals'' that give rise to a band structure. It is thereby possible to analyze the corresponding excitation spectrum engendered by these localized excitations by means of TQC. Special attention is given to the low-energy topological edge mode spectrum. Our work sheds light on the question how the TQC formalism may be generalized to strongly correlated and topologically ordered systems like the Kitaev model.
\end{abstract}

\maketitle

\section{Introduction}\label{sec:Introduction}

In the past ten years the classification of topological phases in condensed matter physics protected by symmetry has seen a tremendous amount of progress, both for non-interacting~\cite{Kru, Bra, Po, Elc} and interacting systems~\cite{Son, Sha, Wang, Che, Ye, Aas, Cheng}. In non-interacting systems, topologically non-trivial phases (like Chern insulators) are usually accompanied by a Wannier obstruction, referring to the fact that the Wannier functions of the occupied set of bands do not behave like exponentially localized orbitals if they also have to respect the symmetries of the system by construction. This interplay between locality and band topology is most intuitively captured by the topological quantum chemistry (TQC) formalism~\cite{Bra, Can}. Using TQC, one infers whether a band insulator or superconductor is topological, depending on the transformation behavior of Bloch states at high-symmetry points in reciprocal space under symmetry.

In contrast, the classification of interacting phases is a more complicated matter. Non-degenerate and gapped topological quantum phases (so-called ``symmetry-protected topological'' (SPT) phases) as well as topologically ordered (so-called ``symmetry-enriched topological'' (SET) phases) are classified using sophisticated mathematical concepts, such as group cohomology, cobordism theory and higher-form symmetries~\cite{Luo}. As the ground states and excitations of interacting systems are not given by products of single-particle Bloch states (in particular in the case of topologically ordered states or spin Hamiltonians), the low energy physics may not be described adequately by band structures anymore. Therefore, the issue of classifying interacting topological phases also poses the question of how to probe the low energy spectrum and in particular the characteristic edge modes experimentally, as these are not given by the usual single-particle excitations studied, for instance, via angular resolved photoemission (ARPES) or scanning tunneling spectroscopy (STS) experiments.

There have been a number of attempts to bridge the methodological gap between the TQC formalism and the existing classification schemes for interacting topological phases. While it is possible to apply TQC to single-particle Green's functions of interacting systems using the topological Hamiltonian framework~\cite{Les}, giving up the restriction of only considering single-particle orbitals seems to be the next step to describe more exotic interacting topological phases~\cite{Ira}. For example, if the interacting state can be qualitatively described by a product state of localized, entangled dimers, which is the case for some SPT phases, then the creation operators of these dimers provide a natural generalization of orbitals and allow for a TQC analysis~\cite{Sol}. More generally, if one can define a set of exponentially localized operators that show the same transformation behavior under symmetry as single-particle states~\cite{Her}, the applicability of TQC appears viable. These operators don't have to be fermionic or describe electronic degrees of freedom at all, as TQC can also be applied to bosons, such as phonons or magnons~\cite{Xu, Kar}.

In this paper, we investigate to what extent the ansatz of generalizing orbitals is applicable to the Kitaev model~\cite{Kit}, as it is a pivotal example of a strongly correlated, topological quantum spin liquid (QSL) and rose to prominence, as its ground state is exactly solvable and due to the fact that it may find its experimental realization in strongly frustrated materials such as ${\alpha}$-RuCl${}_3$~\cite{plumb_PRB2014,johnson_PRB2015,winter2017_review,rousochatzakis2023_review,Bro}. At the same time, we dedicate special attention to the analysis of whether these localized operators accurately describe the characteristic low energy spectrum. This is of special relevance, as the characteristic fractionalization of excitations in QSLs also affects the chiral edge modes of the Kitaev model. After addressing the problem of defining a suitable set of localized operators, we successfully apply a TQC analysis to the excitation spectrum engendered by these localized operators and briefly address, whether they can help in identifying topological edge modes in dynamical structure factors, accessible via experiments like neutron scattering.

This paper is structured as follows: In Sec.~\ref{sec:first_section} we provide a self-contained introduction to the honeycomb Kitaev model, examining the nature of its excitation spectrum as a QSL. Section~\ref{sec:second_section} is devoted to the introduction of localized spin excitations. We study to what extent these can form a set of orthonormal ``orbitals'' and the excitation spectrum that they engender, particularly with respect to edge states. In Sec.~\ref{sec:third_section} we expound how our choice of localized operators can form band representations, serving as the basis for the TQC analysis of the excitation spectrum. Finally, in Sec.~\ref{sec:Conclusion} we conclude our findings and discuss future directions.

\section{The honeycomb Kitaev model}\label{sec:first_section}

  \begin{figure}[ht]
 	\centering
 	\includegraphics[width=1.0\linewidth]{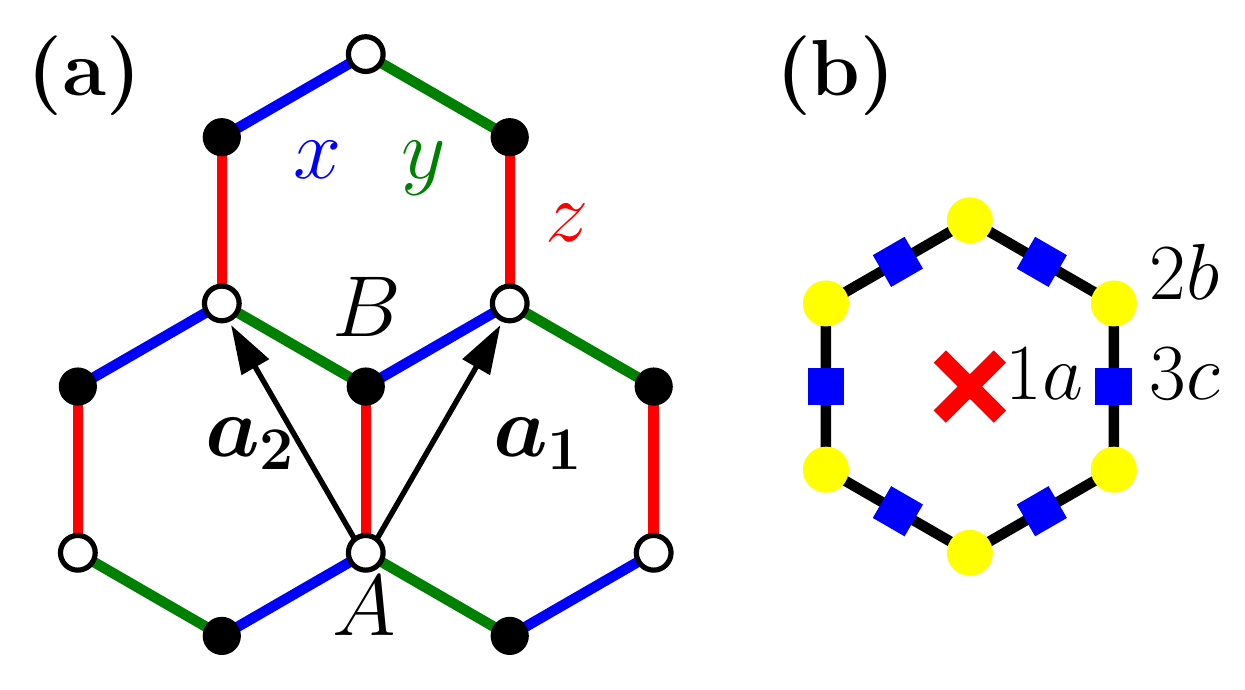}
 	\caption{(a) Bonds, sublattices and unit vectors of the Kitaev model on the honeycomb lattice. Colors denote the three different bond types $x$, $y$ and $z$. (b) Maximal Wyckoff positions of the hexagonal lattice.}
 	\label{fig:kitaev_honeycomb}
 \end{figure}

The spin-1/2 Kitaev model on the honeycomb lattice is defined as~\cite{Kit}
\begin{equation}\label{Eq:Kitaev_honeycomb_spin_model}
\begin{aligned}[b]
    \hat{H} = - \displaystyle\sum_{\vec{R}} & \left\lbrace J_x \sigma_{A, \vec{R} + \vec{a}_1}^{x} \sigma_{B, \vec{R}}^{x} + J_y \sigma_{A, \vec{R} + \vec{a}_2}^{y} \sigma_{B, \vec{R}}^{y} \right.\\
    &\left. + J_z \sigma_{A, \vec{R}}^{z} \sigma_{B, \vec{R}}^{z} \right\rbrace,
\end{aligned}
\end{equation}
with coupling constants $J_x, J_y$ and $J_z$ along $x$, $y$ or $z$ bonds that describe the interaction between neighboring spins, see Fig.~\ref{fig:kitaev_honeycomb}. The position of each spin is determined by a sublattice ($A$ or $B$) and a unit cell $\vec{R}$. Long range magnetic order is suppressed by exchange frustration with different coupling terms of the form $\sigma^{\alpha} \sigma^{\alpha}$ along $\alpha$-bonds.

Using a fermionization approach, it is possible to rewrite the spin model Eq.~\eqref{Eq:Kitaev_honeycomb_spin_model} in terms of itinerant Majorana fermions subject to an effective $\mathbb{Z}_2$ gauge field. The presence of such an effective field is already implied by the ``plaquette'' or ``flux'' operator
\begin{equation}\label{Eq:Flux_operator}
\begin{aligned}[b]
    \hat{w}_{\vec{R}} = & \sigma_{B, \vec{R}}^{z} \sigma_{A, \vec{R} + \vec{a}_2}^{x} \sigma_{B, \vec{R} + \vec{a}_2}^{y} \\
    & \sigma_{A, \vec{R} + \vec{a}_1 + \vec{a}_2}^{z} \sigma_{B, \vec{R} + \vec{a}_1}^{x} \sigma_{A, \vec{R} + \vec{a}_1}^{y}.
\end{aligned}
\end{equation}
of a honeycomb plaquette. This Hermitian operator squares to identity so that its eigenvalues are given by $+1$ and $-1$. Furthermore, these operators commute with each other and the Hamiltonian. On a lattice with periodic boundary conditions with $N$ unit cells, this leads to $N$ conserved ``fluxes'' $w_{\vec{R}}$, which are the eigenvalues of $\hat{w}_{\vec{R}}$. Note though, that on a torus only $N-1$ fluxes can be chosen independently as the total flux through all the plaquettes must satisfy $\prod_{\vec{R}} w_{\vec{R}} = 1$. This implies in particular that flux excitations can only come in pairs. It is expedient to define the bond operators
\begin{equation}
    \hat{K}_{\vec{R}, \vec{R}'}^{\alpha} = \sigma_{A, \vec{R}}^{\alpha} \sigma_{B, \vec{R}'}^{\alpha},
\end{equation}
if $(A, \vec{R})$ and $(B, \vec{R}')$ are an $\alpha$-bond. Omitting the superscript $\alpha$, we can write the flux through a closed path $\vec{\mathcal{R}} = \vec{R}_{1}, \dots, \vec{R}_{n}, \vec{R}_{1}$ as
\begin{equation}\label{Eq:Wilson_loop}
    \hat{w}(\vec{\mathcal{R}}) = \hat{K}_{\vec{R}_1, \vec{R}_2} \hat{K}_{\vec{R}_3, \vec{R}_2} \dots \hat{K}_{\vec{R}_{n-1}, \vec{R}_n} \hat{K}_{\vec{R}_1, \vec{R}_n},
\end{equation}
which, for reasons to become clear soon, we call Wilson-loop operator. The flux through any contractible loop is determined by the set $\lbrace w_{\vec{R}} \rbrace$, however we can identify two additional degrees of freedom by the eigenvalues of the two Wilson-loops $w(\vec{\mathcal{R}}_{1})$ and $w(\vec{\mathcal{R}}_2)$, where $\vec{\mathcal{R}}_1$ and $\vec{\mathcal{R}}_2$ are two arbitrary but fixed loops that are homeomorphic to the two different non-contractible loops on a torus. All in all we can divide the total Hilbert space $\mathcal{H}$ of dimension $2^{2N}$ into $2^{N+1}$ sectors $\mathcal{H}_{w}$ of dimension $2^{N-1}$ that are labelled by a distinct pattern or ``flux configuration'' $w = \lbrace w_{\vec{R}} \rbrace \cup \lbrace w(\vec{\mathcal{R}}_{1}), w(\vec{\mathcal{R}}_2) \rbrace$. It should be noted that the flux described by the Wilson loops is static and cannot propagate through the lattice.

By comparison with Eq.~\eqref{Eq:Flux_operator} we see that a single spin operator, $\sigma_{\gamma, \vec{R}}^{\alpha}$ (with $\gamma \in \lbrace A, B \rbrace$ and $\alpha =x,y,z$) creates two fluxes at the plaquettes neighboring the $\alpha$-bond. We want to express the spin operators in terms of operators that depend on the component $\alpha$ to describe the commutation behavior of $\sigma_{\gamma, \vec{R}}^{\alpha}$ with the flux operators and a set of operators that describes excitations of the remaining degrees of freedom. We accomplish this by the use of Majorana fermions. We make the following ansatz
\begin{equation}\label{Eq:Pauli_as_majorana}
    \tilde{\sigma}_{\gamma, \vec{R}}^{\alpha} = i b_{\gamma, \vec{R}}^{\alpha} c_{\gamma, \vec{R}},
\end{equation}
where $b_{\gamma, \vec{R}}^{\alpha}$ and $c_{\gamma, \vec{R}}$ obey the algebra of Majorana fermions, meaning that they square to one and otherwise anticommute with each other. We include the ``\~{}''-mark to make a distinction to the original Pauli matrices. Evidently, we have extended the former ``physical'' Hilbert space $\mathcal{H}$ to a new Hilbert space $\tilde{\mathcal{H}}$ with dimension $4^{2N}$, as any given pair of ``real'' Majorana fermions $m_1 = m_1^{\dagger}$, $m_2 = m_2^{\dagger}$ forms one ``complex'' fermion $\hat{a} = (m_1 + i m_2)/2$, with $\lbrace \hat{a}^{\dagger}, a\rbrace = 1, \lbrace \hat{a}, \hat{a} \rbrace = 0$, which spans a Hilbert space of dimension $2$. We therefore define
\begin{equation}
    \hat{D}_{\gamma, \vec{R}} = b^{x}_{\gamma, \vec{R}} b^{y}_{\gamma, \vec{R}} b^{z}_{\gamma, \vec{R}} c_{\gamma, \vec{R}}
\end{equation}
and we stipulate that it has the eigenvalue +1 when acting on physical states $\in \mathcal{H}$ and -1 for at least some $\gamma, \vec{R}$ when acting on unphysical ones $\in \tilde{\mathcal{H}} \setminus \mathcal{H}$. We can do this, as $\hat{D}_{\gamma, \vec{R}}$ commutes with the Pauli matrices and hence with every physical observable. For example, for the case of a single lattice site, if $\hat{D} \ket{\uparrow} = + \ket{\uparrow}$, then $\hat{D} \ket{\downarrow} = \hat{D} S^{-} \ket{\uparrow} = + \ket{\downarrow}$, whereas an odd product of Majorana fermions applied on a physical state would produce a state with eigenvalue $-1$. This leads to the following projection operator
\begin{equation}\label{Eq:Projection_operator}
    \hat{P} = \displaystyle\prod_{\vec{R}, \gamma = A, B} \frac{1 + \hat{D}_{\gamma, \vec{R}}}{2},
\end{equation}
which acts as identity on $\mathcal{H}$ and maps states from $\tilde{\mathcal{H}} \setminus \mathcal{H}$ to zero.

Using Eq.~\eqref{Eq:Pauli_as_majorana}, we rewrite the Hamiltonian in Eq.~\eqref{Eq:Kitaev_honeycomb_spin_model} in the extended Hilbert space $\tilde{\mathcal{H}}$ as
\begin{equation}\label{Eq:Extended_hamiltonian}
\begin{aligned}[b]
    \hat{\tilde{H}} = & \frac{\ci J_x}{2} \displaystyle\sum_{\vec{R}} \hat{u}^{x}_{\vec{R} + \vec{a}_1, \vec{R}} c_{A, \vec{R} + \vec{a}_1} c_{B, \vec{R}} \\
    & + \frac{\ci J_y}{2} \displaystyle\sum_{\vec{R}} \hat{u}^{y}_{\vec{R} + \vec{a}_2, \vec{R}} c_{A, \vec{R} + \vec{a}_2} c_{B, \vec{R}} \\
    & + \frac{\ci J_z}{2} \displaystyle\sum_{\vec{R}} \hat{u}^{z}_{\vec{R}, \vec{R}} c_{A, \vec{R}} c_{B, \vec{R}} + \text{h.c.},
\end{aligned}
\end{equation}
with the link operators
\begin{equation}
    \hat{u}^{\alpha}_{\vec{R}', \vec{R}} = i b_{A, \vec{R}'}^{\alpha} b_{B, \vec{R}}^{\alpha},
\end{equation}
where $(A, \vec{R}')$ and $(B, \vec{R})$ describe nearest neighbor sites. Since they have eigenvalues $u_{\vec{R}, \vec{R}'} = \pm 1$, they formally look like Peierls phases~\cite{Lut} associated to $0$ or $\pi$-fluxes, and we can hence interpret them as stemming from a $\mathbb{Z}_2$ gauge field to which the $c_{\gamma, \vec{R}}$ Majorana fermions couple. Furthermore, omitting the superscript $\alpha$ (which is determined by $\vec{R}'$ and $\vec{R}$), we can reexpress Eq.~\eqref{Eq:Wilson_loop} in terms of the link operators as
\begin{equation}
    \hat{w}(\vec{\mathcal{R}}) = \hat{u}_{\vec{R}_1, \vec{R}_2} \hat{u}_{\vec{R}_3, \vec{R}_2} \dots \hat{u}_{\vec{R}_{n-1}, \vec{R}_n} \hat{u}_{\vec{R}_1, \vec{R}_n},
\end{equation}
so all Wilson loops can be uniquely determined by the set of all link operators $u = \lbrace u_{\vec{R}, \vec{R}'} \rbrace$. It should be noted that even though the link operators commute with each other and the Hamiltonian in Eq.~\eqref{Eq:Extended_hamiltonian}, they anticommute with $\prod_{\gamma, \vec{R}} \hat{D}_{\gamma, \vec{R}}$, so $\hat{D}_{\gamma, \vec{R}}$ can be regarded as a gauge transformation that flips the sign of the link operators on neighboring links.

To diagonalize the Hamiltonian we first pick a specific configuration of the link operators $u$ corresponding to a flux configuration $w$. Eq.~\eqref{Eq:Extended_hamiltonian} can then be treated like a hopping Hamiltonian with the important difference that the $c_{\gamma, \vec{R}}$ operators are Majorana fermions.

Since we focus on topological properties, we are only interested in the ground state of the Kitaev Hamiltonian in the physical Hilbert space $\mathcal{H}$. To find it, we solve the Kitaev model in the extended Hilbert space and we then project the ground state onto the physical Hilbert space. First, we identify a link operator configuration $u$ corresponding to the flux configuration $w$ that minimizes the ground state energy, then we can proceed to diagonalize the now quadratic Hamiltonian. By expressing pairs of two ``real'' Majorana fermions as ``complex'' fermions, eigenstates can then be written with the use of fermionic occupation numbers. There is however an important subtlety due to the projection operator in Eq.~\eqref{Eq:Projection_operator}: since the flux configurations $\lbrace w \rbrace$ account for $2^{N+1}$ degrees of freedom, upon projection into the physical Hilbert space the ``complex'' fermionic modes can only account for $2^{N-1}$ modes instead of $2^{N}$. This apparent paradox can be resolved by understanding that a complex fermion is a linear superposition of Majorana fermions $c_{\gamma, \vec{R}}$ which anticommute with $\prod_{\gamma, \vec{R}} \hat{D}_{\gamma, \vec{R}}$. Therefore, all physical states of the same flux configuration must have the same parity of excited complex fermionic modes. Note that $\sum_{k \text{ even}} \binom{N}{k} = \sum_{k \text{ odd}} \binom{N}{k} = 2^{N-1}$. Physical excitations (meaning excitations between states in the physical Hilbert space) that do not affect the flux configuration must come in pairs of two fermionic modes. This explains the quantum spin liquid character of the Kitaev model: Excitations of flux degrees of freedom as well as of itinerant Majorana modes for a given flux configuration must come in pairs.

\begin{figure}[ht]
 	\centering
 	\includegraphics[width=1.0\linewidth]{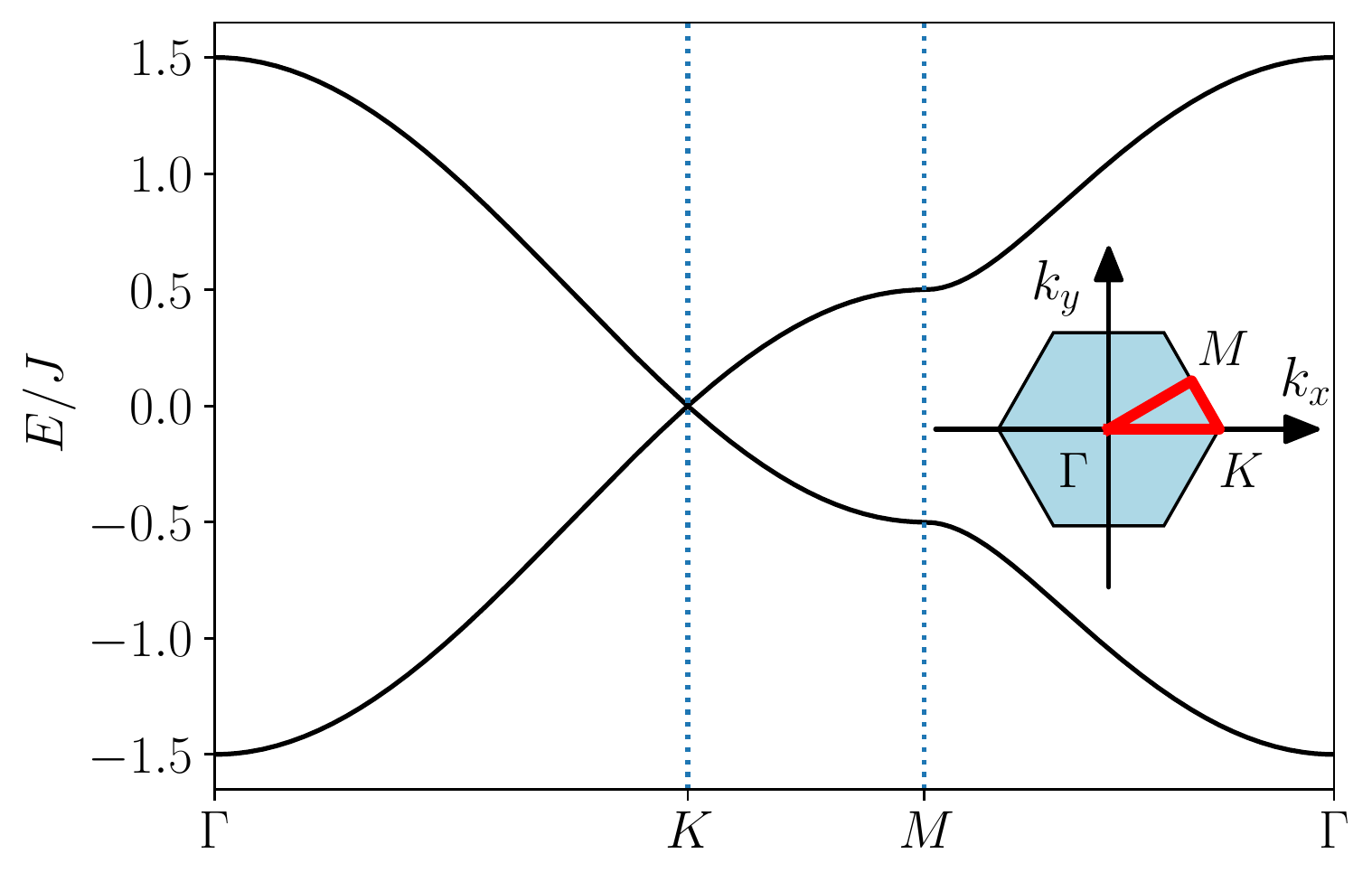}
 	\caption{``Majorana spectrum'' of $\tilde{h}(\vec{k})$ (see Eq.~\eqref{Eq:h_tilde}) along the high-symmetry path for $K = 0$.}
 \label{fig:majorana_spectrum_high_symmetry}
 \end{figure}

According to a theorem from Lieb~\cite{Lie} the ground state of the Kitaev model is given by the flux-free configuration $w_{\vec{R}} = 1$ for all $\vec{R}$ on the infinite plane, see Appendix~\ref{App:first_appendix}. We realize the flux-configuration by setting $u_{\vec{R}, \vec{R}'} = 1$ which we denote as $u = 1$. The Hamiltonian in Eq.~\eqref{Eq:Extended_hamiltonian} then formally looks like that of graphene with nearest neighbor hopping and can straightforwardly be diagonalized. By defining the Fourier-transformation
\begin{equation}
    c_{\gamma, \vec{R}} = \frac{1}{\sqrt{N}} \displaystyle\sum_{\vec{k}} \e^{-i \vec{k} \vec{R}} c_{\gamma, \vec{k}}
\end{equation}
we rewrite the Hamiltonian in the flux-free sector as
\begin{equation}\label{Eq:Graphene_like_Kitaev}
    \hat{\tilde{H}}_{u = 1} = \displaystyle\sum_{\vec{k}} \vec{c}_{\vec{k}}^{\dagger} \begin{pmatrix}
    0 & \frac{\ci}{2} f_{\vec{k}} \\
    - \frac{\ci}{2} f^{*}_{\vec{k}} & 0
    \end{pmatrix} \vec{c}_{\vec{k}},
\end{equation}
with $\vec{c}_{\vec{k}} = (c_{A, \vec{k}}, c_{B, \vec{k}})^{T}$ and $f_{\vec{k}} = J_x \e^{i \vec{k} \vec{a}_1} + J_y \e^{i \vec{k} \vec{a}_2} + J_z$. The spectrum of the matrix in Eq.~\eqref{Eq:Graphene_like_Kitaev} is plotted in Fig.~\ref{fig:majorana_spectrum_high_symmetry}.

   \begin{figure}[ht]
 	\centering
 	\includegraphics[width=1.0\linewidth]{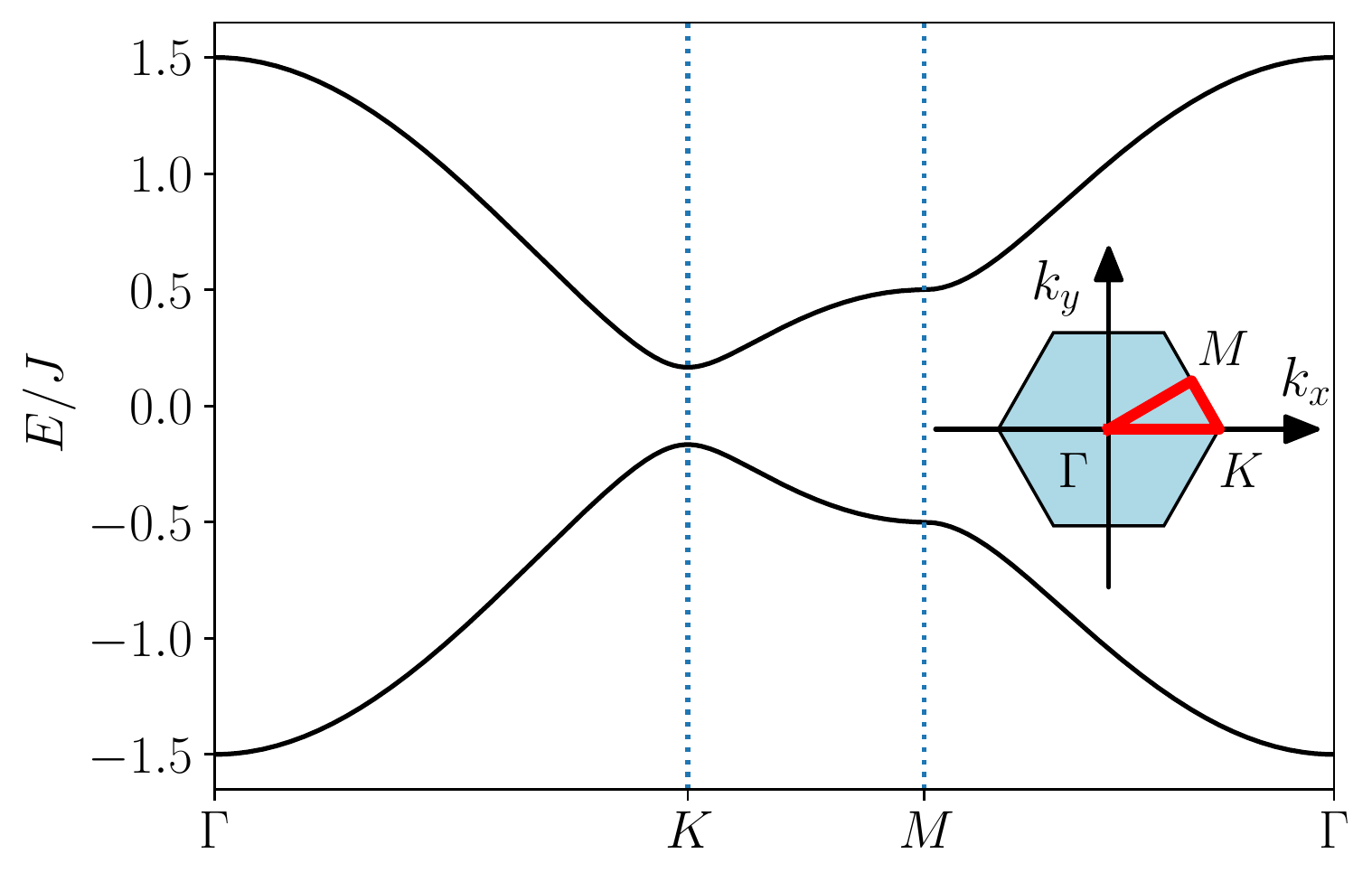}
 	\caption{``Majorana spectrum'' of $\tilde{h}(\vec{k})$ (see Eq.~\eqref{Eq:h_tilde}) along the high-symmetry path for $K = 0.064$.}
 \label{fig:majorana_spectrum_high_symmetry_K}
 \end{figure}

In the periodic table of topological invariants~\cite{Ryu}, the Hamiltonian in the extended Hilbert space $\tilde{\mathcal{H}}$ in Eq.~\eqref{Eq:Extended_hamiltonian} for any fixed configuration $u$ belongs to the symmetry class BDI, which in two dimensions does not permit topological invariants solely protected by time reversal and particle hole symmetry. We can however break time reversal symmetry by adding the term
\begin{equation}\label{Eq:Kitaev_magnetic_field}
    \begin{aligned}
        \hat{H}_{K} =  - K \displaystyle\sum_{\vec{R}} & \left\lbrace \sigma_{B, \vec{R} + \vec{a}_2}^{x} \sigma_{A, \vec{R} + \vec{a}_1 + \vec{a}_2}^{z} \sigma_{B, \vec{R} + \vec{a}_1}^{y} \right.\\
        & + \sigma_{A, \vec{R} + \vec{a}_1 + \vec{a}_2}^{y} \sigma_{B, \vec{R} + \vec{a}_1}^{x} \sigma_{A, \vec{R} + \vec{a}_1}^{z} \\
        & + \sigma_{B, \vec{R} + \vec{a}_1}^{z} \sigma_{A, \vec{R} + \vec{a}_1}^{y} \sigma_{B, \vec{R}}^{x} \\
        & + \sigma_{A, \vec{R} + \vec{a}_1}^{x} \sigma_{B, \vec{R}}^{z} \sigma_{A, \vec{R} + \vec{a}_2}^{y}\\
        & + \sigma_{B, \vec{R}}^{y} \sigma_{A, \vec{R} + \vec{a}_2}^{x} \sigma_{B, \vec{R} + \vec{a}_2}^{z} \\
        & \left. + \sigma_{A, \vec{R} + \vec{a}_2}^{z} \sigma_{B, \vec{R} + \vec{a}_2}^{y} \sigma_{A, \vec{R} + \vec{a}_1 + \vec{a}_2}^{x}\right\rbrace,
    \end{aligned}
\end{equation}
which can be obtained by expanding a magnetic field term that couples to the spins like $\vec{\sigma} \vec{h}$, with $K = h_x h_y h_z/J^2$~\cite{Kit, Zha2}. In the extended Hilbert space in the flux free $u=1$ sector, and then setting all $\hat{D}_{\gamma, \vec{R}} = 1$ (denoted as $D = 1$) as we are ultimately only interested in the physical Hilbert space, it yields
\begin{equation}
\begin{aligned}[b]
    & \hat{\tilde{H}}_{K, u = 1, D = 1} \\
    = & \frac{\ci}{2} K \displaystyle\sum_{\vec{R}} \left\lbrace c_{B, \vec{R} + \vec{a}_2} c_{B, \vec{R} + \vec{a}_1} + c_{A, \vec{R} + \vec{a}_1 + \vec{a}_2} c_{A, \vec{R} + \vec{a}_1} \right. \\
    &  + c_{B, \vec{R} + \vec{a}_1} c_{B, \vec{R}} + c_{A, \vec{R} + \vec{a}_1} c_{A, \vec{R} + \vec{a}_2}\\
    & \left.+ c_{B, \vec{R}} c_{B, \vec{R} + \vec{a}_2} + c_{A, \vec{R} + \vec{a}_2} c_{A, \vec{R} + \vec{a}_1 + \vec{a}_2} \right\rbrace + \text{h.c.}     
\end{aligned}
\end{equation}
The Hamiltonian now reads
\begin{equation}\label{Eq:h_tilde}
\begin{aligned}[b]
    \hat{\tilde{H}}_{u = 1} + \hat{\tilde{H}}_{K, u=1, D=1} = \displaystyle\sum_{\vec{k}} \vec{c}^{\dagger}_{\vec{k}} \tilde{h}(\vec{k}) \vec{c}_{\vec{k}} \\
    = \displaystyle\sum_{\vec{k}} \vec{c}^{\dagger}_{\vec{k}} \begin{pmatrix}
    \Delta_{\vec{k}} & \frac{\ci}{2} f_{\vec{k}} \\
    - \frac{\ci}{2} f^{*}_{\vec{k}} & -\Delta_{\vec{k}}
    \end{pmatrix} \vec{c}_{\vec{k}},
\end{aligned}
\end{equation}
with $\Delta(\vec{k}) = K (\sin \vec{k} \vec{a}_1 - \sin \vec{k} (\vec{a}_1 - \vec{a}_2) - \sin \vec{k} \vec{a}_2)$. The spectrum of $\tilde{h}(\vec{k})$ for nonzero $K$ is displayed in Fig.~\ref{fig:majorana_spectrum_high_symmetry_K}. Equation~\eqref{Eq:h_tilde} formally looks like the Haldane model~\cite{Hal}. Its symmetry class changes to D, which does permit integer invariants in two dimensions in the form of Chern numbers of $\tilde{h}(\vec{k})$. It should be noted that Lieb's theorem relies on reflection symmetry, which is broken for non-zero $K$. For sufficiently small $K$ however, we can expect that the ground state will still lie in the flux-free sector (see Appendix~\ref{App:first_appendix}). 

 One can further transform the Fourier transformed Majorana fermions via
\begin{equation}
    \begin{pmatrix}
        c_{A, \vec{k}} \\
        c_{B, \vec{k}}
    \end{pmatrix} = \sqrt{2} \begin{pmatrix}
        \sin \varphi_{\vec{k}} \e^{i \phi_{\vec{k}}} & - \cos \varphi_{\vec{k}} \e^{i \phi_{\vec{k}}} \\
        \cos \varphi_{\vec{k}} & \sin \varphi_{\vec{k}}
    \end{pmatrix} \begin{pmatrix}
        a_{\vec{k}} \\
        a_{-\vec{k}}^{\dagger}
    \end{pmatrix}
\end{equation}
and
\begin{equation}
\begin{gathered}[b]
    \sin^2 \varphi_{\vec{k}} = \frac{1}{2} + \frac{2 \Delta_{\vec{k}}}{\epsilon_{\vec{k}}}, \quad \cos^2 \varphi_{\vec{k}} = \frac{1}{2} - \frac{2 \Delta_{\vec{k}}}{\epsilon_{\vec{k}}}, \\
    e^{i \phi_{k}} = i \frac{|f_{\vec{k}}|}{f_{\vec{k}}},
    \end{gathered}
\end{equation}
where $\epsilon_{\vec{k}} = 4 \sqrt{\Delta_{\vec{k}}^2 + |\vec{f}_{\vec{k}}|^2/4}$ (note the factor of 4 compared to Eq.~\eqref{Eq:h_tilde}) into complex fermions $a_{\vec{k}}$, with $\lbrace a_{\vec{k}}, a_{\vec{k}'} \rbrace = 0$ and $\lbrace a_{\vec{k}}, a_{\vec{k}}^{\dagger} \rbrace = \delta_{\vec{k}, \vec{k}'}$. We finally get
\begin{equation}\label{Eq:diagonal_kitaev}
    \hat{H} =\displaystyle\sum_{\vec{k}} \epsilon_{\vec{k}} (a_{\vec{k}}^{\dagger} a_{\vec{k}} - 1/2),
\end{equation}
so the ground state in the extended Hilbert space is given by the vacuum state of $a_{\vec{k}}$ fermions. It should be noted that on a torus geometry in the flux-free sector only states with an odd parity of $a_{\vec{k}}$ fermions turn out to be physical states~\cite{Ped}. Therefore, in the physical ground state, one fermionic mode has to be excited. This has consequences for the many-body spectrum, but will be negligible for any observable in the thermodynamic limit.

  \begin{figure}[ht]
 	\centering
 	\includegraphics[width=1.0\linewidth]{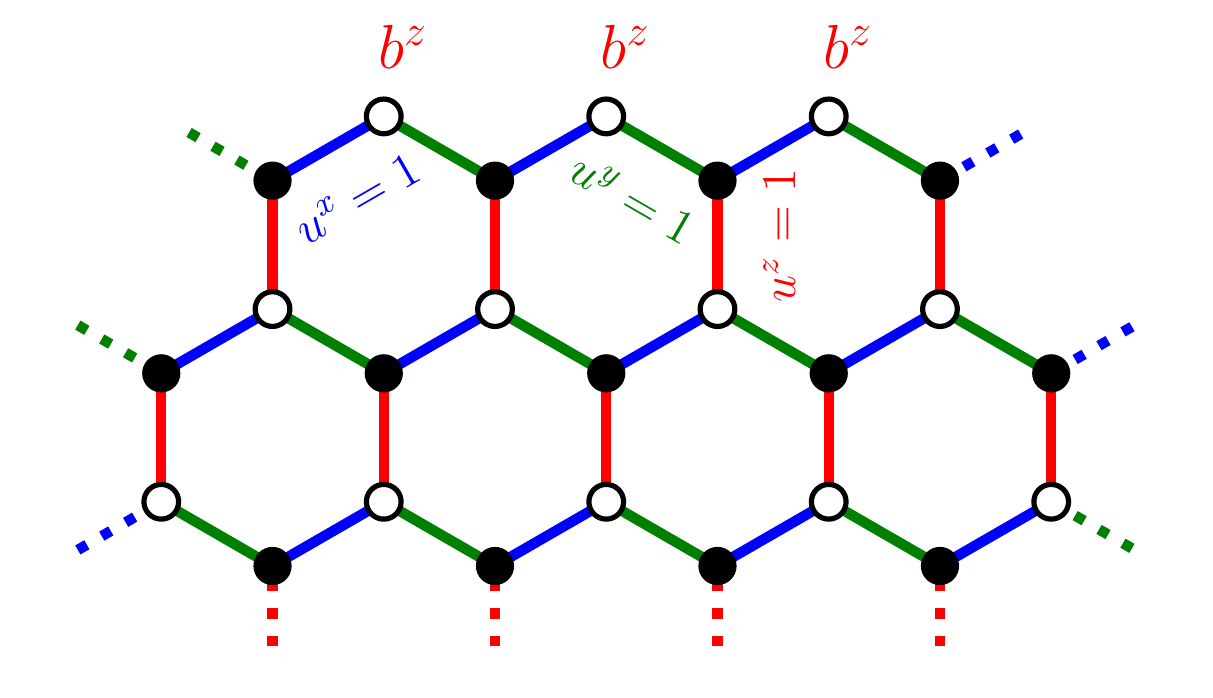}
 	\caption{Zigzag edge of the Kitaev model. The $b^z$ Majoranas at the $A$-sites of the edge do not have another $b^z$ Majorana fermion to couple to.}
 	\label{fig:kitaev_zigzag}
 \end{figure}

   \begin{figure}[ht]
 	\centering
 	\includegraphics[width=1.0\linewidth]{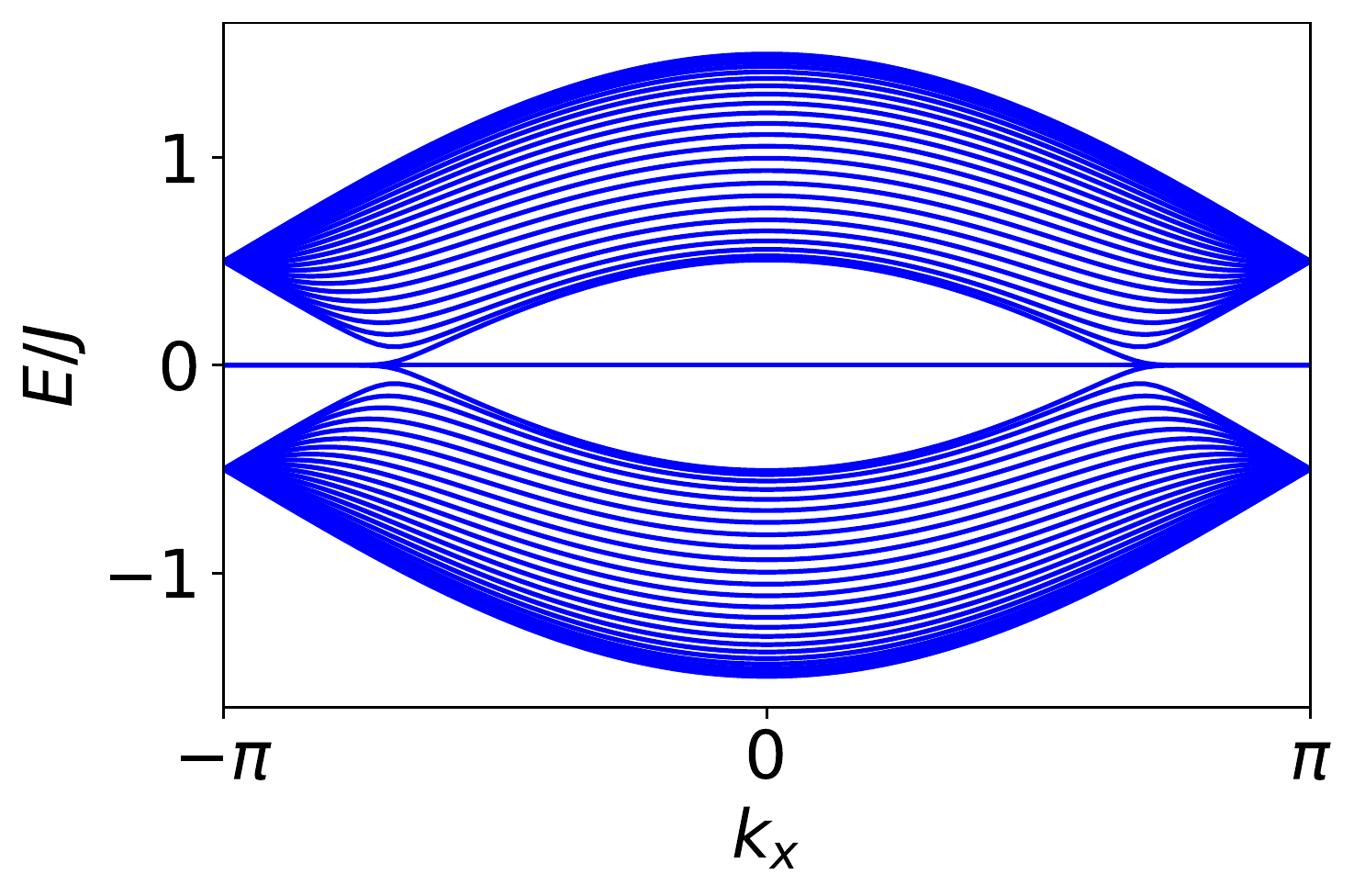}
 	\caption{``Majorana spectrum'' of $\tilde{h}(k_x)$, see Eq.~\eqref{Eq:hamiltonian_matrix_open_boundaries} for $K = 0$.}
 	\label{fig:Kitaev_edge_modes_gapless}
 \end{figure}

   \begin{figure}[ht]
 	\centering
 	\includegraphics[width=1.0\linewidth]{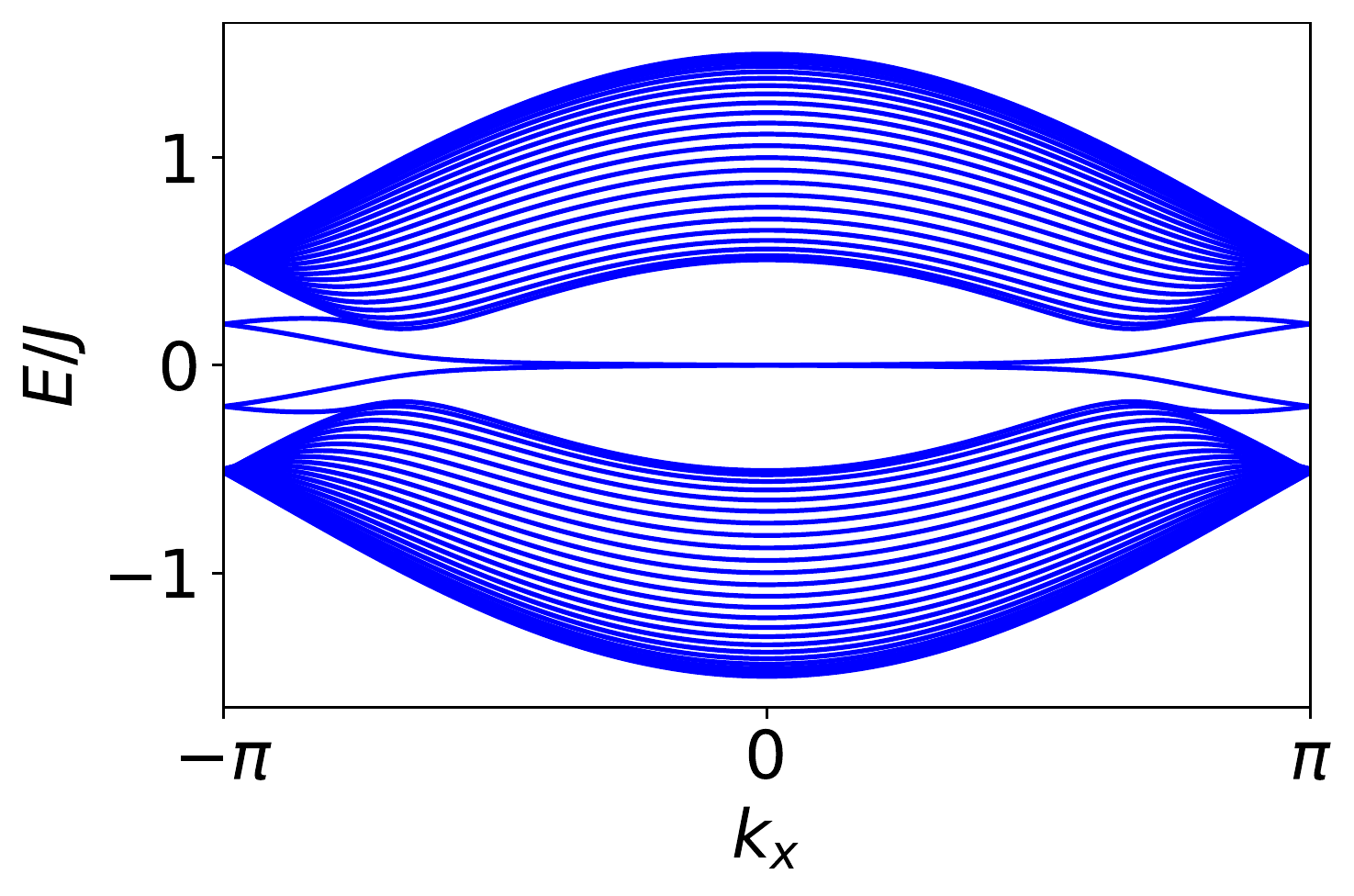}
 	\caption{``Majorana spectrum'' of $\tilde{h}(k_x)$, see Eq.~\eqref{Eq:hamiltonian_matrix_open_boundaries} for $K = 0.064$.}
 	\label{fig:Kitaev_edge_modes_gapped}
 \end{figure}

If the triangle inequalities
\begin{equation}\label{Eq:triangle_inequalities}
    |J_{\alpha}| \leq |J_{\beta}| + |J_{\gamma}|
\end{equation}
for $\alpha \neq \beta \neq \gamma \neq \alpha \in \lbrace x,y,z \rbrace$ are satisfied, then the Kitaev model without TRI-breaking in Eq.~\eqref{Eq:Graphene_like_Kitaev} will have two gapless bands, that can gap out with any non-zero $K$. Each band then has a nonzero Chern number. If the triangle inequalities are violated, the Majorana bands will always be trivially gapped, independent of the value of $K$. A non-zero Chern number is accompanied by gapless edge modes on a geometry with open boundary conditions. Figures~\ref{fig:Kitaev_edge_modes_gapless} and \ref{fig:Kitaev_edge_modes_gapped} show the Majorana spectrum for $K = 0$ and $K = 0.064$, with chiral edge modes in the later case, on a zigzag geometry, see Fig.~\ref{fig:kitaev_zigzag}. Note that at the edges there are uncoupled zero energy $b^z$ Majorana modes for $K = 0$ that can, due to the TRI breaking term, couple to the itinerant $c_{\gamma, \vec{R}}$ modes and hence become itinerant themselves, see Appendix~\ref{App:third_appendix} for further details. As a final note we mention that independently of the band topology of the Majorana bands the Hamiltonian always realizes topologically non-trivial QSL phases, albeit they may belong to different topological equivalence classes~\cite{Kit}.

\section{Itinerant excitations}\label{sec:second_section}

The ground state of the Kitaev model on the infinite plane is given by the vacuum state $\ket{0}$ of the $a_{\vec{k}}$ operators. As noted before however, single particle excitations of $a_{\vec{k}}$ modes will not result in physical modes, as these do not commute with the product of gauge operators $\hat{D}_{\gamma, \vec{R}}$. Instead, applying (products of) spin operators on the ground state will always lead to a new state in the physical Hilbert space and can hence be used to engender excitations. This also has the following conceptual advantage. The addition of further terms such as the off-diagonal Gamma interaction, important for the modeling of $\alpha$-RuCl${}_{3}$~\cite{Win}, to the Hamiltonian may lead to the vacuum state not being the ground state anymore. Such a term actually appears in the expansion of the TRI-breaking magnetic field, which is usually neglected as it contains four-fermion terms and hence not being of immediate relevance for the low energy physics. It may also be possible that the $a_{\vec{k}}$ modes do not describe the low energy physics anymore at all. This will be the case at very large magnetic fields that destroy the QSL-phase. Nevertheless, applying spin operators on the ground state of a spin Hamiltonian is always well defined and will engender a spectrum of excitations, making our ansatz also realizable in numerical techniques such as exact diagonalization or density matrix renormalization group.

In the following, (products of) localized spin operators $\hat{O}_{\vec{R}}^{\mu}$ will serve as our starting point for employing the language of orbitals to the Kitaev model. We will ultimately try to expand the Hamiltonian in terms of the Fourier-transformed $\hat{O}^{\mu}_{\vec{q}} = 1/\sqrt{N} \sum_{\vec{R}} \e^{\ci \vec{q} \vec{R}} \hat{O}^{\mu}_{\vec{R}}$, which has a formal analogy to magnon physics~\cite{Kar}. Since these terms will later be identified as playing the same role as localized orbitals, we will call them ``spin orbitals'' from now on. We focus on operators that do not create any flux excitations for two reasons. First of all, let the operators $\hat{O}^{\mu}_{\vec{R}}$ and $\hat{O}^{\nu}_{\vec{R}'}$ create fluxes at $\vec{R}$ and $\vec{R}'$ respectively. Then, as states with different flux configurations are orthogonal to each other, it is straightforward to show that there will not be any non-trivial $\vec{q}$-space geometry of states spanned by $\hat{O}^{\mu}_{\vec{q}} \ket{0}$, when expanding operators commuting with the flux operators, such as the Hamiltonian. Secondly, the topologically interesting edge modes consist of excitations of itinerant Majorana fermions, not of the static flux excitations.

There are, in principle, infinitely many different ways to define $\hat{O}^{\mu}_{\vec{R}}$. Since single Pauli matrix terms are ruled out for periodic boundary conditions as they always create flux excitations, we consider only terms of two or three neighboring Pauli matrices:
\begin{equation}\label{Eq:orbitals}
    \begin{aligned}[b]
        \hat{O}^{x}_{\vec{R}} = & \sigma_{A, \vec{R}}^{x} \sigma^{x}_{B, \vec{R} - \vec{a}_{1}}, \\
        \hat{O}^{y}_{\vec{R}} = & \sigma_{A, \vec{R}}^{y} \sigma^{y}_{B, \vec{R} - \vec{a}_2}, \\
        \hat{O}^{z}_{\vec{R}} = & \sigma_{A, \vec{R}}^{z} \sigma^{z}_{B, \vec{R}}, \\
        \hat{O}^{A_1}_{\vec{R}} = & \sigma_{A, \vec{R}}^{z} \sigma_{B, \vec{R}}^{y} \sigma_{A, \vec{R} + \vec{a}_1}^{x}, \\
        \hat{O}^{A_2}_{\vec{R}} = & \sigma_{A, \vec{R} + \vec{a}_2}^{y} \sigma_{B, \vec{R}}^{x} \sigma_{A, \vec{R}}^{z}, \\
        \hat{O}^{A_3}_{\vec{R}} = & \sigma_{A, \vec{R} + \vec{a}_1}^{x} \sigma_{B, \vec{R}}^{z} \sigma_{A, \vec{R} + \vec{a}_2}^{y}, \\
        \hat{O}^{B_1}_{\vec{R}} = & \sigma_{B, \vec{R}}^{z} \sigma_{A, \vec{R}}^{y} \sigma_{B, \vec{R} - \vec{a}_1}^{x}, \\
        \hat{O}^{B_2}_{\vec{R}} = & \sigma_{B, \vec{R} - \vec{a}_2}^{y} \sigma_{A, \vec{R}}^{x} \sigma_{B, \vec{R}}^{z}, \\
        \hat{O}^{B_3}_{\vec{R}} = & \sigma_{B, \vec{R} - \vec{a}_1}^{x} \sigma_{A, \vec{R}}^{z} \sigma_{B, \vec{R} - \vec{a}_2}^{y}.
    \end{aligned}
\end{equation}
All terms with three or less Pauli matrices must be localized around some $\vec{R}$ if they are not to create any flux excitations, justifying our choice in Eq.~\eqref{Eq:orbitals}. We are only interested in their action on the ground state, so link and gauge operators are set to one and, in normal order representation, terms containing one or more annihilation operators vanish. We can then always write the above terms acting on the ground state in the form
\begin{equation}\label{Eq:orbitals_on_ground_state}
    \hat{O}^{\mu}_{\vec{q}} \to \frac{1}{\sqrt{N}} \displaystyle\sum_{\vec{k}} \left( \Omega^{\mu}_{\vec{k}, \vec{q}} a_{\vec{k}}^{\dagger} a_{-\vec{k}-\vec{q}}^{\dagger} + \tilde{\Omega}^{\mu}_{\vec{k}, \vec{q}} \delta_{\vec{q}, \vec{0}} \right),
\end{equation}
see Appendix~\ref{App:fourth_1_appendix}. These spin orbitals acting on the ground state hence can be used to exhaust a part of the low-energy Hilbert space spanned by $a_{\vec{k}}^{\dagger} a_{\vec{k}'}^{\dagger} \ket{0}$. Note that it is not possible to write the terms
\begin{equation}\label{Eq:interaction_orbitals}
    \begin{aligned}[b]
        \hat{O}^{3A}_{\vec{R}} & = \sigma_{A, \vec{R}}^{z} \sigma_{A, \vec{R} + \vec{a}_1}^{x} \sigma_{A, \vec{R} + \vec{a}_2}^{y} \\
        \hat{O}^{3B}_{\vec{R}} & = \sigma_{B, \vec{R}}^{z} \sigma_{B, \vec{R} - \vec{a}_1}^{x} \sigma_{B, \vec{R} - \vec{a}_2}^{y}        
    \end{aligned}
\end{equation}
in the form of Eq.~\eqref{Eq:orbitals_on_ground_state} even though they commute with the flux operators, because they also contain terms with four creation operators. Focusing on low energy excitations, we will therefore not consider these in the following.

We have defined a possible basis of spin orbitals, but the issue of orthonormality still needs to be addressed. It can readily be checked that $\braket{0|(\hat{O}_{\vec{q}}^{\mu})^{\dagger} \hat{O}^{\nu}_{\vec{q}'}|0} \propto \delta_{\vec{q}, \vec{q}'}$, however for the same $\vec{q}$ the spin orbitals are in general not orthonormal or may even be linearly dependent: for example, for isotropic exchange couplings and $K = 0$ one finds
\begin{equation}\label{Eq:linear_dependence}
    \displaystyle \lim_{\substack{\vec{q} \to \vec{0} \\ \vec{q} \neq \vec{0}}} \left( \hat{O}_{\vec{q}}^{x} + \hat{O}_{\vec{q}}^{y} + \hat{O}^{z}_{\vec{q}}\right) \ket{0} = 0.
\end{equation}
We define the overlap matrix
\begin{equation}\label{Eq:overlap_matrix}
    S_{\vec{q}}^{\mu \nu} = \braket{0|\left( \hat{O}_{\vec{q}}^{\mu} \right)^{\dagger} \hat{O}_{\vec{q}}^{\nu}|0}
\end{equation}
so that we obtain orthonormal spin orbitals
\begin{equation}
    \hat{\bar{O}}^{\mu}_{\vec{q}} = \left( S_{\vec{q}}^{-\frac{1}{2}} \right)_{\alpha \mu} \hat{O}^{\alpha}_{\vec{q}}.
\end{equation}
In this basis it is possible to expand the Hamiltonian (note that $S_{\vec{q}}$ is hermitian):
\begin{equation}\label{Eq:hamiltonian_expansion}
\begin{aligned}[b]
    H^{\mu \nu}_{\vec{q}} = & \braket{0|\left( \hat{\bar{O}}^{\mu}_{\vec{q}} \right)^{\dagger} (\hat{H} + \hat{H}_{K} - E_0) \hat{\bar{O}}^{\nu}_{\vec{q}} |0} \\
    = & \displaystyle\sum_{\alpha, \beta} \left( S_{\vec{q}}^{-\frac{1}{2}} \right)_{\mu \alpha} \left( S_{\vec{q}}^{-\frac{1}{2}} \right)_{\beta \nu} \frac{1}{N} \displaystyle\sum_{\vec{k}} (\epsilon_{\vec{k}} + \epsilon_{\vec{k} + \vec{q}}) \\
    & \cdot \left( \left( \Omega^{\alpha}_{\vec{k}, \vec{q}} \right)^{*} \Omega^{\beta}_{\vec{k}, \vec{q}} - \left( \Omega^{\alpha}_{\vec{k}, \vec{q}} \right)^{*} \Omega^{\beta}_{-\vec{k}-\vec{q}, \vec{q}} \right).
\end{aligned}
\end{equation}
Here, the sums over $\alpha$ and $\beta$ go over the number of spin orbitals under consideration and $E_0$ is the ground state energy.

\begin{figure}[ht]
 	\centering
 	\includegraphics[width=1.0\linewidth]{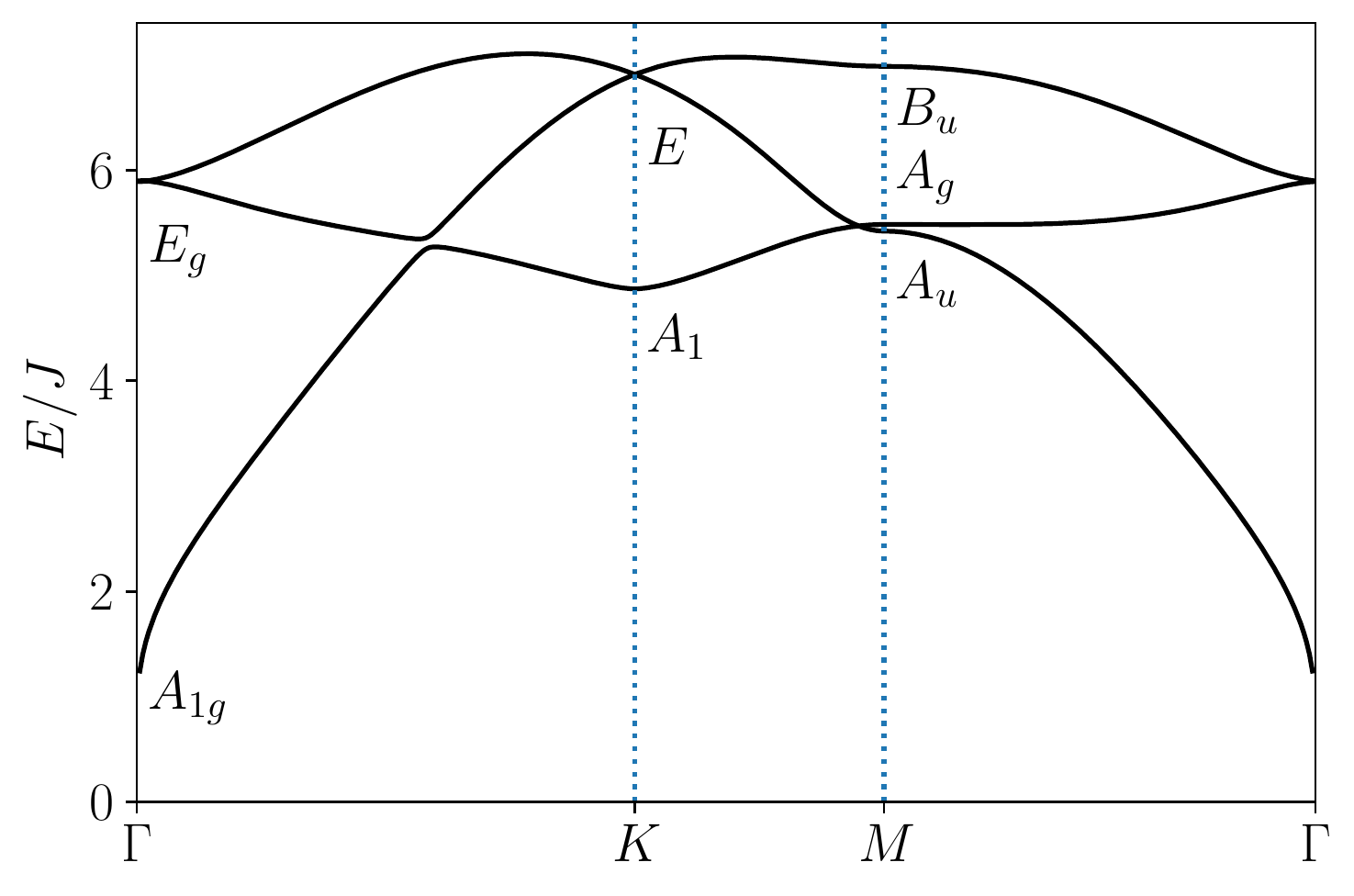}
 	\caption{Eigenvalue spectrum of $H_{\vec{q}}$ for isotropic exchange coupling $J = 1$ and $K = 0$. Eigenvalues at high-symmetry points are labeled according to the irreps of the little cogroups of how the Bloch functions transform, see Sec.~\ref{sec:third_section}.}
 	\label{fig:H_exp_3_orb_K_0.0_J_1}
 \end{figure}

 \begin{figure}[ht]
 	\centering
 	\includegraphics[width=1.0\linewidth]{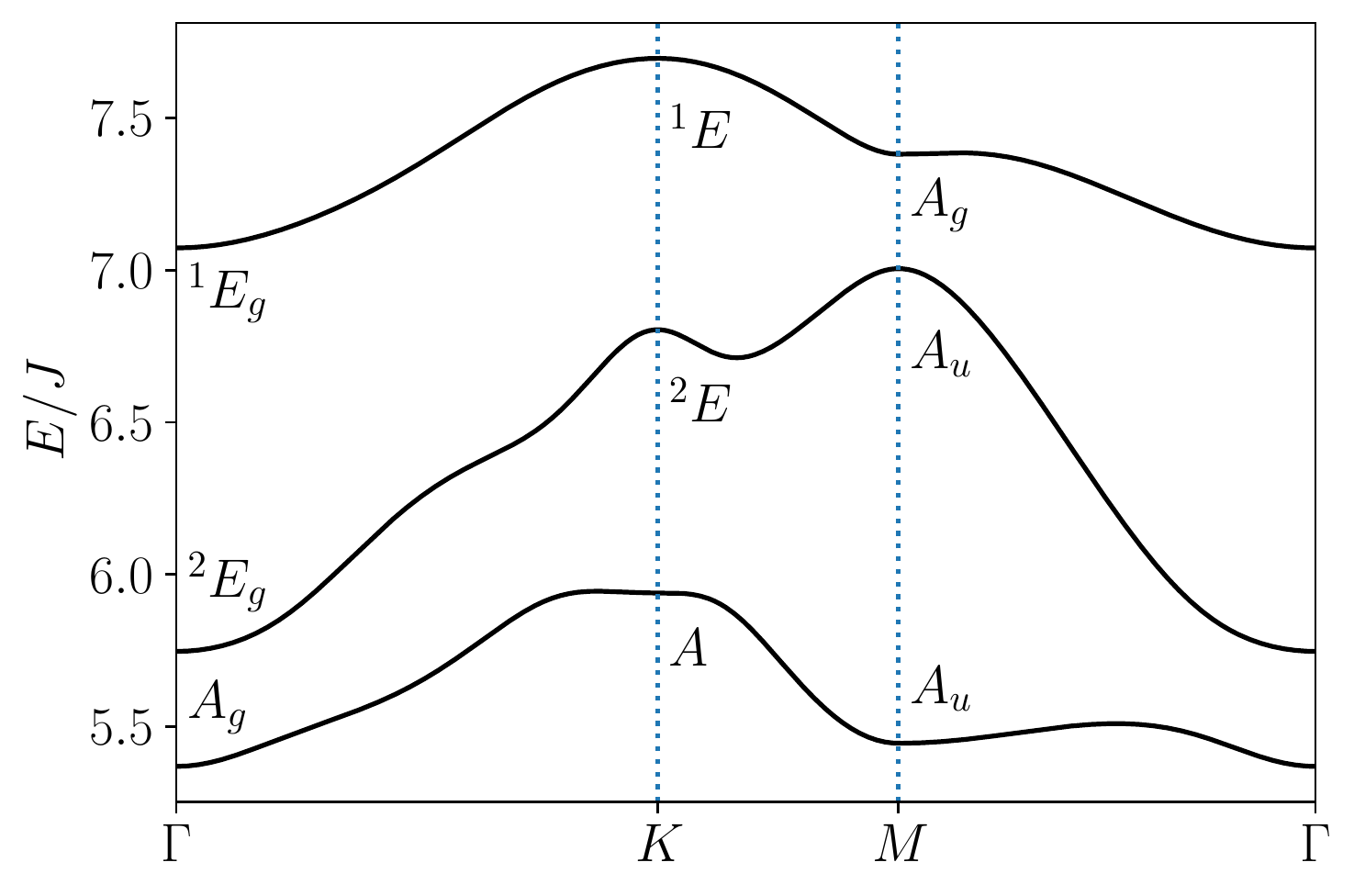}
 	\caption{Eigenvalue spectrum of $H_{\vec{q}}$ for isotropic exchange coupling $J = 1$ and $K = 0.2$. Eigenvalues at high-symmetry points are labeled according to the irreps of the little co-groups of how the Bloch functions transform, see Sec.~\ref{sec:third_section}.}
 	\label{fig:H_exp_3_orb_K_0.2_J_1}
 \end{figure}

  \begin{figure}[ht]
 	\centering
 	\includegraphics[width=1.0\linewidth]{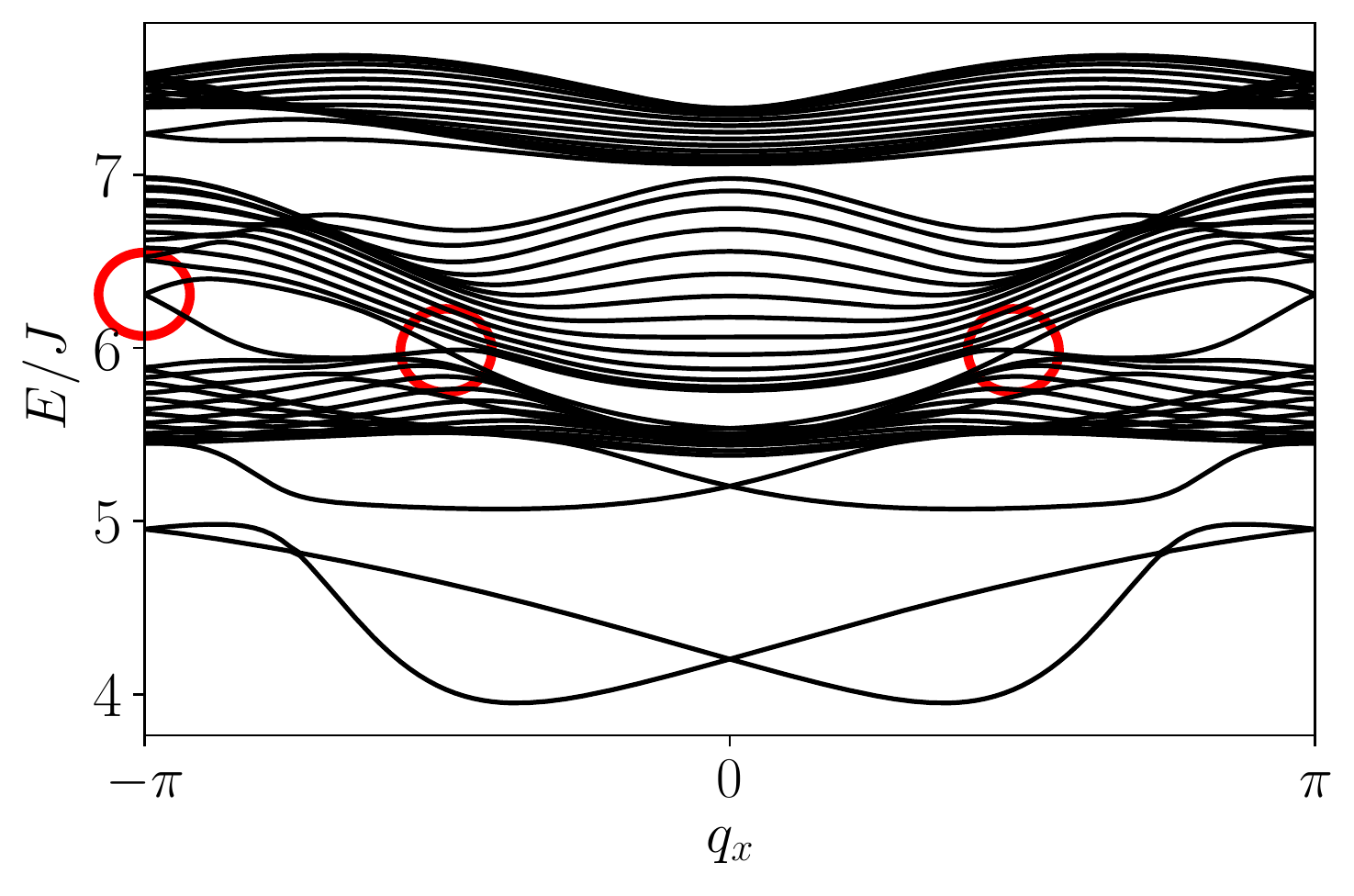}
 	\caption{Eigenvalue spectrum of $H_{q_x}$ (see Eq.~\eqref{Eq:hamiltonian_expansion_open_boundaries}) for isotropic exchange coupling $J = 1$ and $K = 0.2$. Chiral edge modes between the two energetically lowest bands are indicated by red circles. Only the $\hat{O}_j^{x, i}$, $\hat{O}_j^{y, i}$ and $\hat{O}_j^{z, i}$ spin orbitals are considered.}
 	\label{fig:H_exp_edge_1_orb_K_0.2_J_1}
 \end{figure}

   \begin{figure}[ht]
 	\centering
 	\includegraphics[width=1.0\linewidth]{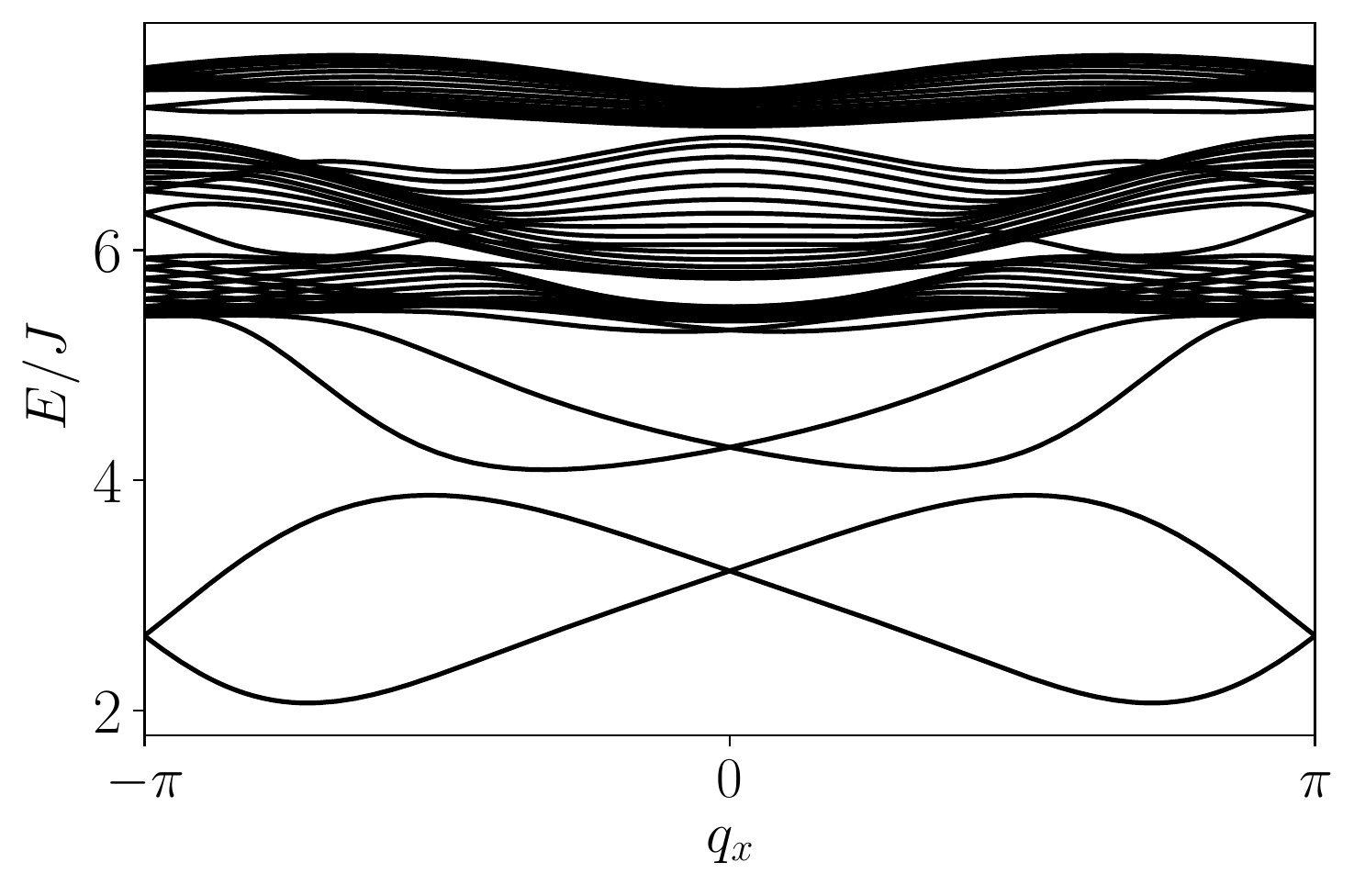}
 	\caption{Eigenvalue spectrum of $H_{q_x}$ (see Eq.~\eqref{Eq:hamiltonian_expansion_open_boundaries}) for isotropic exchange coupling $J = 1$ and $K = 0.2$. In addition to the $\hat{O}_j^{x, i}$, $\hat{O}_j^{y, i}$ and $\hat{O}_j^{z, i}$ spin orbitals, here also the two single-spin orbitals in Eq.~\eqref{Eq:sigma_z_spin_orbitals} are considered.}
 	\label{fig:H_exp_edge_1_with_b_z_K_0.2_J_1}
 \end{figure}

For the sake of simplicity we start with excitations from the three spin orbitals $\hat{O}^{x}_{\vec{R}}, \hat{O}^{y}_{\vec{R}}$ and $\hat{O}^{z}_{\vec{R}}$ containing two Pauli matrices (e.g. relevant for Raman scattering~\cite{Kno}, where the Raman vertex operator has the same form as the first three terms in Eq.~\eqref{Eq:orbitals}). Figures~\ref{fig:H_exp_3_orb_K_0.0_J_1} and \ref{fig:H_exp_3_orb_K_0.2_J_1} show the spectrum of Eq.~\eqref{Eq:hamiltonian_expansion} for isotropic exchange couplings $J_{\alpha} = J = 1$ and $K = 0$ and $K = 0.2$~\footnote{Note that we had to exclude a data point in the direct vicinity of the $\Gamma$-point due to extremely bad convergence. Issues due to poor convergence occured also in some plots of Appendix~\ref{App:fourth_appendix}. For the data of the plots of Eq.~\eqref{Eq:hamiltonian_expansion} and Eq.~\eqref{Eq:hamiltonian_expansion_open_boundaries}, see Supplemental Material~\cite{supplement}.}. For the case of broken TRI the degeneracies in Fig.~\ref{fig:H_exp_3_orb_K_0.0_J_1} (in particular between the $K$- and $M$-point) are lifted and the spectrum becomes gapped. The symmetry indicators of the two lowest bands indicate a Chern number $C = 3 \text{ mod } 6$ (see Sec.~\ref{sec:third_section}). On a stripe geometry (see Fig.~\ref{fig:kitaev_zigzag}) this indicates three chiral edge modes, as seen in Fig.~\ref{fig:H_exp_edge_1_orb_K_0.2_J_1}. Very notably though, some isolated modes also protrude below the bulk band spectrum. Their nature as edge modes becomes apparent when including the single Pauli matrix spin orbitals at the vertices of the zigzag edge that do not induce any flux excitations as their $b_z$ modes do not have any partner to couple to (see Eq.~\eqref{Eq:sigma_z_spin_orbitals}). Figure~\ref{fig:H_exp_edge_1_with_b_z_K_0.2_J_1} shows a distinct lowering of the energy of the isolated modes due to the fact that the almost dispersionless $b^z$ modes are excited by the two additional spin orbitals. We stress that these modes are energetically lower than bulk excitations and hence indicate stable modes in the exact spectrum of $\hat{H}$ below that of its bulk states.

Upon inclusion of additional spin orbitals (for calculations also considering the spin orbitals with three Pauli matrices, see Appendix~\ref{App:fourth_appendix}) more and more of the low energy Hilbert space can be exhausted in the spectrum of $H_{\vec{q}}$. It should be noted though that even for all spin orbitals in Eq.~\eqref{Eq:orbitals} the overall excitation energies are far above the flux gap of about $\approx 0.24J$~\cite{Kit} so that much of the spectral weight will be high in energy, in contrast to single spin excitations accessible via neutron scattering that have the bulk of their spectral weight in energies below $1J$ even though they excite flux degrees of freedom~\cite{Kno2}. Hence, many higher energy modes from the remaining dispersion will be excited by local products of spin excitations, in particular when attempting to probe edge modes. This is in contrast to thermal transport at very low temperatures that only excites the itinerant edge modes. We can conclude that these modes must have their origin in long chains of spin excitations to be energetically low enough. The topology of the spectrum of $H_{\vec{q}}$ and $H_{q_x}$ (see Eq.~\eqref{Eq:hamiltonian_expansion_open_boundaries}) is nevertheless useful as it indicates the presence of localized excitations with spectral weight predominantly at the edge. It would be interesting for future studies to inquire how much of the spectral weight of these modes has to be localized entirely at the edge (that is, if two fermionic modes are excited that are both localized entirely at the edge or whether edge modes tend to come with bulk excitations).


\section{Band representations of local spin excitations}\label{sec:third_section}

The basis of TQC is to study space group representations that are induced by exponentially localized orbitals obeying all crystalline symmetries. Let an exponentially localized orbital at some point in real space belonging to a Wyckoff position $\text{Wyck}$ transform under a representation $\rho$ of its site-symmetry group. Then one can define a representation $\rho_{G}$ of a space group $G$ by $\rho_{G} = (\bar{\rho} \uparrow G)_{\text{Wyck}}$, where $\bar{\rho}$ is a point group representation equivalent to $\rho$ of a point group isomorphic to the site-symmetry group. If Wannier functions of some band of Bloch states can also be exponentially localized while obeying all symmetries, then they too must transform under some band representation.
At high symmetry points $\vec{k}$, Bloch functions transform under some irrep (irreducible representation) of the little group $G_{\vec{k}}$. The set of these irreps, called a quasi-band representation~\cite{Can}, serves as a symmetry indicator, as only specific sets of quasi-band representations can be associated to band representations; all other possible quasi-band representations hence indicate a Wannier obstruction due to nontrivial topology.

In the fully symmetric case, the space group is given by $P\bar{3}1m1'$. The $\hat{O}^{x}$, $\hat{O}^{y}$ and $\hat{O}^{z}$ orbitals are then localized at the $3c$ Wyckoff position, while the remaining spin orbitals in Eq.~\eqref{Eq:orbitals} are at the $2b$ position\footnote{It should be noted that we are technically dealing with a layer group which we treat as a wallpaper group due to the lack of a $z$-axis. As a consequence, the notation for Wyckoff positions differs from that one used in the Bilbao Crystallographic Server~\cite{Aro}. In addition, when working with the ``topological quantum chemistry''~\cite{Elc} tool, one has to keep in mind the different dimension: Whereas a 3D material with the magnetic space group 162.77 (the case with non-zero K) and orbitals in the 3f Wyckoff position (which is equivalent to the 3c Wyckoff position in the corresponding wallpaper group) does not allow a gapped band structure of different elementary band representations, this is possible in 2D due to the absence of an A, H and L point in the BZ.}, see Fig.~\ref{fig:kitaev_honeycomb}. It is straightforward to find the representations of the site-symmetry groups for the spin orbitals at $3c$, which is equivalent to $A_g$, and at $2b$, which is equivalent to $A_2 \oplus E$; the induced band representations are then given by $(A_g \uparrow G)_{3c}$ and $(A_2 \oplus E \uparrow G)_{2b}$ and for different symmetries displayed in Table~\ref{tab:space_groups} analogously. At high-symmetry points $\vec{k}$, we label Bloch states by irreducible representations of the little cogroup $\tilde{G}_{\vec{k}}$ (which is the quotient group of $G_{\vec{k}}$ divided by the translation group) with the projection operator
\begin{equation}
    \hat{P}_{\vec{k}}^{\mu} = \frac{d^{\mu}_{\vec{k}}}{N_{\vec{k}}} \displaystyle\sum_{g \in G_{\vec{k}}} \chi_{\vec{k}}^{\mu}(g)^{*} T_{\vec{k}}(g),
\end{equation}
where $\mu$ is the irrep of dimension $d^{\mu}_{\vec{k}}$ and character $\chi^{\mu}_{\vec{k}}$ in question, $N_{\vec{k}}$ is the order of the little cogroup $\tilde{G}_{\vec{k}}$, and $T_{\vec{k}}$ is the representation under which the orbitals transform, given in Appendix~\ref{App:fourth_2_appendix}. For further details, see Ref.~\cite{Can2} and see Ref.~\cite{Aro2} for the character tables of the different point groups.

Symmetry indicators provide more information than solely the presence of a Wannier obstruction. In the case of rotation symmetry, one can determine the Chern number $C^{\nu}$ of the $\nu$-th band of the spectrum of $H_{\vec{q}}$ via~\cite{Fan}
\begin{equation}
    \e^{\ci \pi C^{\nu} / 3} = T_{\Gamma}^{\nu}(S_{6}^{+}) T_{K}^{\nu}(C_{3}^{+}) T_{M}^{\nu}(I),
\end{equation}
where $T^{\nu}_{\vec{k}}$ is the irreducible representation of a Bloch state at some high-symmetry point $\vec{k}$. For the isotropic case (with TRI-breaking field) they are given by $D_{3d}$ ($C_{3i}$) at the $\Gamma$ point, $C_{3v}$ ($C_3$) at the $K$ point and $C_{2h}$ ($C_i$) at the $M$ point. One readily finds a Chern number of 3 modulo 6 for the two energetically lowest bands and a Chern number of $0$ modulo 6 for the remaining band for the parameters and spin orbitals chosen for Fig.~\ref{fig:H_exp_3_orb_K_0.2_J_1}.


As a final note we mention that in some scenarios (e.g. in the case of a staggered TRI-breaking magnetic field) the current analysis should be extended to include magnetic space groups~\cite{Elc}. For future studies it is also possible to add additional spin-exchange terms to the Hamiltonian so that the ground state is in a different flux configuration~\cite{Zha}. This allows in principle for an extension of the TQC analysis to include magnetic translation operators~\cite{Fang}.

\section{Conclusion and Outlook}\label{sec:Conclusion}

In this work we have studied a possible extension of the topological quantum chemistry (TQC) formalism to strongly correlated systems such as the Kitaev honeycomb model. The expansion of the Hamiltonian in terms of localized spin operators serving as orbitals leads to a band structure which can be analyzed by means of TQC. A Wannier obstruction due to nontrivial topology was identified via symmetry indicators, leading to edge modes in an open boundary geometry. Remarkably though, there appear also localized edge states energetically below bulk excitations, in stark contrast to edge modes in topological magnon systems.

For future study, an extension of our proposed framework to Green's functions come to mind~\cite{Sol}, in particular in the context of the topological Hamiltonian approach~\cite{Les, Wan}. This remains a challenge in particular when it comes to the interpretation of such an ansatz, as we are not dealing with some product state of localized dimers~\cite{Sol, Her}. In addition, the feasibility of creating a Green's function formalism with operators similar to the Hubbard X operators remains an open question.

There are two restrictions that were used in the choice of orbitals. As indicated in Sec.~\ref{sec:second_section}, a band structure originating from single Pauli matrices at the $2b$ Wyckoff position has to be topologically trivial, as the flux patterns stemming from $\sigma^{x}$, $\sigma^{y}$ and $\sigma^{z}$ Pauli matrices are different from each other. It can be made topologically nontrivial though, once hopping processes between flux degrees of freedom are allowed~\cite{Joy}: The present ansatz has the advantage that it can be generalized to include these hopping terms straightforwardly. In addition, it is also possible to consider localized operators in an extended Hilbert space, such as the itinerant Majorana modes $c_{\gamma, \vec{R}}$ or Abrikosov fermions~\cite{Bur} and Schwinger bosons~\cite{Kos}.

Finally, we raise the question of the stability of the edge modes. As was mentioned in Sec.~\ref{sec:second_section}, it is also possible to include an ``interaction term'' in the Kitaev model (see Eq.~\eqref{Eq:interaction_orbitals}), which appears naturally when expanding the TRI-breaking magnetic field given by $\vec{h} \vec{\sigma}$~\cite{Kit}. When expressing Majorana fermions in terms of complex fermions, this leads to terms in the Hamiltonian violating particle number conservation. For Chern insulators, it is known that charge conservation plays an important role in protecting gapless edge states~\cite{Qi}, so it stands to reason that the edge mode spectrum may be fragile to interaction, similarly to topological Heisenberg systems~\cite{Hab}, although this has not been demonstrated numerically yet. We leave this question and its implications for transport to future investigation.

\begin{acknowledgments}

We wish to thank Juan L. Ma\~{n}es, Francesco Ferrari and David A. S. Kaib
 for stimulating discussions. This work was supported by the Deutsche Forschungsgemeinschaft (DFG) through QUAST-FOR5249 - 449872909 (project TP4). M.G.V. and M.G.D. acknowledge the Spanish Ministerio de Ciencia e Innovación (PID2022-142008NB-I00), the Canada Excellence Research Chairs Program for Topological Quantum Matter and  the Ministry for Digital Transformation and of Civil Service of the Spanish Government through the QUANTUM ENIA project call - Quantum Spain project, and by the European Union through the Recovery, Transformation and Resilience Plan - NextGenerationEU within the framework of the Digital Spain 2026 Agenda. M.G.D. acknowledges financial support from Government of the Basque Country through the pre-doctoral fellowship PRE 2023 2 0024.

\end{acknowledgments}

\appendix

\section{Lieb's flux theorem}\label{App:first_appendix}

Lieb showed for a general class of fermions on bipartite lattices which flux pattern maximizes the grand-canonical partition function or subsequently minimizes the ground state energy at zero temperature~\cite{Lie}. The ground state energy is always minimized at half filling which is the reason, why the theorem also holds for quadratic Majorana Hamiltonians: any Hamiltonian of the form
\begin{equation}\label{Eq:app_majorana_hamiltonian}
    \hat{H} = \frac{\ci}{4} \vec{c}^{T} A \vec{c},
\end{equation}
with a real, skew symmetric $2N \times 2N$ matrix $A$ with eigenvalues $\lbrace \pm  \ci \epsilon_j \rbrace$ and some Majorana modes $\vec{c} = (c_1, c_2, \dots)^{T}$ can be rewritten as
\begin{equation}\label{Eq:app_diagonal_hamiltonian}
    \hat{H} = \displaystyle\sum_{j} \epsilon_j (a_j^{\dagger} a_j - 1/2),
\end{equation}
with complex fermions $a_j$, $\lbrace a_i^{\dagger}, a_j \rbrace = \delta_{ij}, \lbrace a_i, a_j \rbrace = 0$, where the sum only goes over the non-negative half of the eigenvalues~\cite{Kit}. We minimize the ground state energy by minimizing $- \sum_{j} \epsilon_j$. If $\ci A$ corresponds to one of the bipartite lattices considered in~\cite{Lie}, then Lieb's theorem will also hold for Eq.~\eqref{Eq:app_majorana_hamiltonian}. It should be noted that this correspondence only works for Hamiltonians that are quadratic in Majorana fermions: Whereas Lieb's theorem also encompasses some interaction terms, it does not automatically hold for Hamiltonians with quartic Majorana terms, since they cannot be written in the form of Eq.~\eqref{Eq:app_diagonal_hamiltonian}. For a different perspective on the applicability of Lieb's theorem, specifically for Majorana systems, see~\cite{Per}.

In case of the honeycomb lattice with nearest neighbor hopping, the ground state energy is minimized by having zero flux per plaquette. The proof relies on reflection symmetry, which is present in the Kitaev Hamiltonian in Eq.~\eqref{Eq:Kitaev_honeycomb_spin_model} if at least two of the three coupling constants are equal (see also Appendix~\ref{App:second_appendix}), but which is violated upon including a TRI breaking term like in Eq.~\eqref{Eq:Kitaev_magnetic_field}. While phase transitions to different flux patterns have been theoretically predicted as a function of a TRI breaking magnetic field~\cite{Chu, Li} or even mirror symmetry preserving next-next-nearest neighbor Majorana hopping~\cite{Zha}, for sufficiently small fields the flux gap stays finite and stable~\cite{Mot}, so in the rest of the paper we will only work within the flux free sector\footnote{We finally note the subtlety that in the flux-free sector on the torus the ground state energy of a physical state cannot in general be given by $-\sum_{j} \epsilon_j$, since the parity of complex modes $a_{j}$ must be odd~\cite{Ped}. While this is not expected to change the physics in the thermodynamic limit, it may in principle have an effect on the flux pattern in a finite sized cluster.}.

\section{Symmetries of the Kitaev model}\label{App:second_appendix}

Depending on whether the bonds $J_x, J_y$ and $J_z$ are isotropic and whether the TRI breaking field with non-zero $K$ is present influences under which symmetries the Hamiltonian $\hat{H} +
 \hat{H}_{K}$ stays invariant. Symmetries consists of translations given by integer multiples of the unit vectors $\vec{a}_1$ and $\vec{a}_2$, crystalline symmetries and crystalline symmetries multiplied with time reversal. Table~\ref{tab:space_groups} provides a list of the symmetry elements.

\begin{table*}[]
\begin{ruledtabular}\begin{tabular}{cccc}
$K = 0 ?$ & Exchange couplings & Space group & Symmetry elements (without translations)  \\ \hline \hline
Yes & $J_x = J_y = J_z$ & $P \bar{3} 1m  1'$ (162.74) & $\lbrace E, C_3^{\pm}, C_{2i}', I, S_{6}^{\pm}, \sigma_{vi} \rbrace \otimes \lbrace E, \hat{\mathcal{T}} \rbrace$ \\ \hline
Yes & $J_x \neq J_y = J_z$ & $C2/m1' (12.59)$ & $\lbrace E, C_{23}', I, \sigma_{v2} \rbrace \otimes \lbrace E, \hat{\mathcal{T}} \rbrace$ \\ \hline
Yes & $J_x \neq J_y \neq J_z$ & $P\bar{1} 1' (2.5)$ & $\lbrace E, I \rbrace \otimes \lbrace E, \hat{\mathcal{T}} \rbrace$ \\ \hline
No & $J_x = J_y = J_z$ & $P \bar{3} 1m' $ (162.77) & $\lbrace E, C_{3}^{\pm}, I, S_6^{\pm}, \hat{\mathcal{T}} C_{2i}', \hat{\mathcal{T}} \sigma_{vi} \rbrace$ \\ \hline
No & $J_x \neq J_y = J_z$ & $C2'/m'$ (12.62) & $\lbrace E, I, \hat{\mathcal{T}} C_{23}', \hat{\mathcal{T}} \sigma_{v2} \rbrace$ \\ \hline
No & $J_x \neq J_y \neq J_z$ & $P\bar{1}$ (2.4) & $\lbrace E, I \rbrace$
\end{tabular}\end{ruledtabular}
\caption{Space groups and coset representatives with respect to the translation group of the honeycomb Kitaev model depending on the presence of a TRI breaking term $K$ and for different exchange couplings.}
\label{tab:space_groups}
\end{table*}

  \begin{figure}[ht]
 	\centering
 	\includegraphics[width=0.9\linewidth]{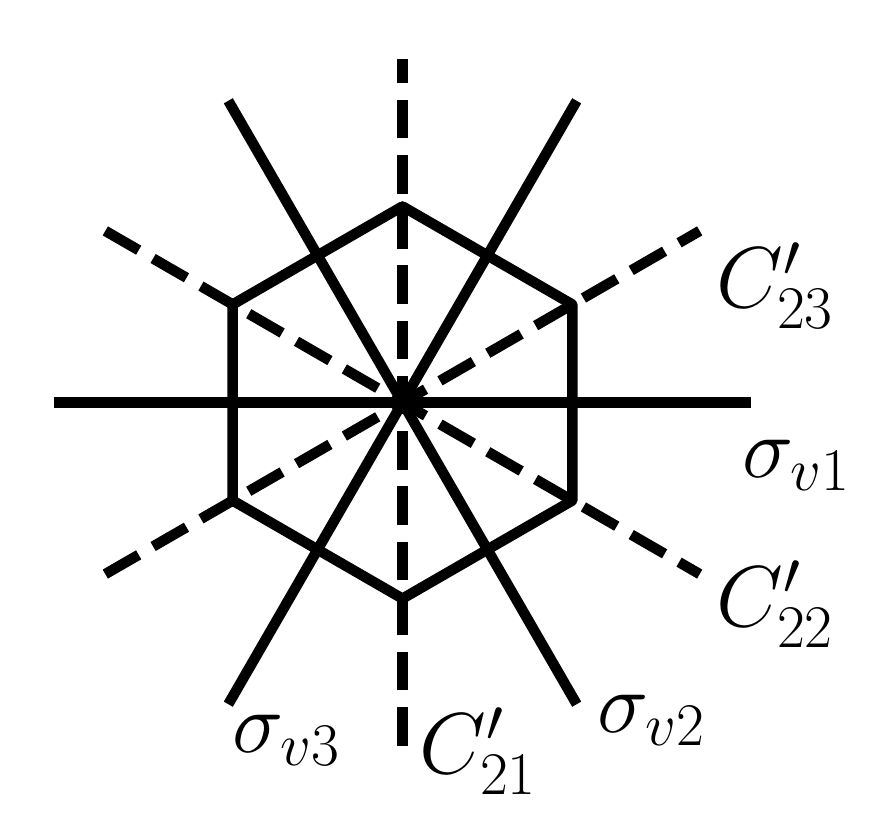}
 	\caption{Mirror plane and secondary rotation symmetries of the honeycomb Kitaev model. The principle rotation axis (not shown) is at the center of the plaquette.}
 	\label{fig:Kitaev_symmetries}
 \end{figure}

The action of the different symmetry elements are displayed in Fig.~\ref{fig:Kitaev_symmetries} and Table~\ref{tab:symmetries}. Due to strong spin-orbit coupling, which is the cause of the exchange frustration, the transformation behavior of the spin operators must be considered, too.

\begin{table*}[]
\begin{ruledtabular}\begin{tabular}{cccc}
Symmetries & $(n_1', n_2')$ & $(\vec{\delta}_{A}', \vec{\delta}_{B}')$ & $(x',y',z')$ \\ \hline \hline
$E$ & $(n_1, n_2)$ & $(\vec{\delta}_{A}, \vec{\delta}_{B})$ & $(x,y,z)$ \\ \hline
$\hat{\mathcal{T}}$ & $(n_1, n_2)$ & $(\vec{\delta}_{A}, \vec{\delta}_{B})$ & $(-x,-y,-z)$ \\ \hline
$C_3^{+}$ & $(-n_1-n_2, n_1)$ & $(\vec{\delta}_{A}-\vec{a}_1, \vec{\delta}_{B}-2\vec{a}_1)$ & $(y,z,x)$ \\ \hline
$C_3^{-}$ & $(n_2, -n_1-n_2)$ & $(\vec{\delta}_{A}-\vec{a}_2, \vec{\delta}_{B}-2\vec{a}_2)$ & $(z,x,y)$ \\ \hline
$C_{21}'$ & $(n_2, n_1)$ & $(\vec{\delta}_{A}, \vec{\delta}_{B})$ & $(-y,-x,-z)$ \\ \hline
$C_{22}'$ & $(-n_1-n_2, n_2)$ & $(\vec{\delta}_{A}-\vec{a}_1, \vec{\delta}_{B}-2\vec{a}_1)$ & $(-z,-y,-x)$ \\ \hline
$C_{23}'$ & $(n_1, -n_1-n_2)$ & $(\vec{\delta}_{A} -\vec{a}_2, \vec{\delta}_{B}-2\vec{a}_2)$ & $(-x,-z,-y)$ \\ \hline
$I$ & $(-n_1, -n_2)$ & $(\vec{\delta}_{B}-\vec{a}_1-\vec{a}_2, \vec{\delta}_{A}-\vec{a}_1-\vec{a}_2)$ & $(x,y,z)$ \\ \hline
$S_{6}^{+}$ & $(-n_2, n_1 + n_2)$ & $(\vec{\delta}_{B}-\vec{a}_1, \vec{\delta}_{A}-\vec{a}_1+\vec{a}_2)$ & $(z,x,y)$ \\ \hline
$S_{6}^{-}$ & $(n_1+n_2, -n_1)$ & $(\vec{\delta}_{B}-\vec{a}_2, \vec{\delta}_{A}+\vec{a}_1-\vec{a}_2)$ & $(y,z,x)$ \\ \hline
$\sigma_{v1}$ & $(-n_2, -n_1)$ & $(\vec{\delta}_{B}-\vec{a}_1-\vec{a}_2, \vec{\delta}_{A}-\vec{a}_1 -\vec{a}_2)$ & $(-y,-x,-z)$ \\ \hline
$\sigma_{v2}$ & $(-n_1, n_1+n_2)$ & $(\vec{\delta}_{B}-\vec{a}_1, \vec{\delta}_{A}-\vec{a}_1 + \vec{a}_2$) & $(-x,-z,-y)$ \\ \hline
$\sigma_{v3}$ & $(n_1+n_2, -n_2)$ & $(\vec{\delta}_{B} -\vec{a}_2, \vec{\delta}_{A}+\vec{a}_1 - \vec{a}_{2})$ & $(-z,-y,-x)$
\end{tabular}\end{ruledtabular}
\caption{Action of the symmetries on the unit cell $\vec{R} = n_1 \vec{a}_1 + n_2 \vec{a}_2 \mapsto n_1' \vec{a}_1 + n_2' \vec{a}_2$, sublattices $(\vec{\delta}_{A}, \vec{\delta}_{B}) \mapsto (\vec{\delta}_{A}', \vec{\delta}_{B}')$ and spin operators $(S_x, S_y, S_z) \equiv (x,y,z) \mapsto (x',y',z')$.}
\label{tab:symmetries}
\end{table*}

\section{Edge modes of the Kitaev model}\label{App:third_appendix}

\subsection{Diagonalization}\label{App:third_1_appendix}

On an infinitely extended stripe geometry in $\vec{x}$ direction with zigzag edges, see Fig.~\ref{fig:kitaev_zigzag}, the Kitaev Hamiltonian in Eq.~\eqref{Eq:Kitaev_honeycomb_spin_model} becomes
\begin{equation}
\begin{aligned}
    \hat{H} = & - J_x \displaystyle\sum_{j \in \Lambda_{x}} \displaystyle\sum_{i = 1}^{N_A} \sigma_{A_i, j}^{x} \sigma_{B_{i-1}, j}^{x} \\
    & - J_y \displaystyle\sum_{j \in \Lambda_{x}} \displaystyle\sum_{i = 1}^{N_A} \sigma_{A_i, j-1}^{y} \sigma_{B_{i-1}, j}^{y} \\
    & - J_z \displaystyle\sum_{j \in \Lambda_{x}} \displaystyle\sum_{i = 1}^{N_A - 1} \sigma_{A_i, j}^{z} \sigma_{B_{i}, j}^{z},
\end{aligned}
\end{equation}
where $\Lambda_x$ is the number of sites in $\vec{x}$-direction for a given row and sublattice, $N_A$ is the number of rows for each sublattice $\vec{\delta}_{A}, \vec{\delta}_{B}$ and the position of the sites is defined by
\begin{equation}
    \begin{aligned}
    (A_i, j): \vec{\delta}_A + i \vec{a}_1 + j (\vec{a}_1 - \vec{a}_2), & \quad i \in \lbrace 1, \dots, N_A \rbrace\\
    (B_i, j): \vec{\delta}_B + i \vec{a}_1 + j (\vec{a}_1 - \vec{a}_2), & \quad i \in \lbrace 0, \dots, N_A - 1 \rbrace.
    \end{aligned}
\end{equation}

Likewise, for the TRI breaking term in Eq.~\eqref{Eq:Kitaev_magnetic_field} we find
\begin{equation}
\begin{aligned}[b]
    \hat{H}_{K} = -K \displaystyle\sum_{j \in \Lambda_x} & \left( \displaystyle\sum_{i = 1}^{N_A} \sigma_{B_{i-1}, j}^{x} \sigma_{A_i, j}^{z} \sigma_{B_{i-1}, j+1}^{y} \right. \\
    & + \displaystyle\sum_{i = 1}^{N_A - 1} \sigma_{A_{i+1}, j-1}^{y} \sigma_{B_i, j}^{x} \sigma_{A_i, j}^{z} \\
    & + \displaystyle\sum_{i = 1}^{N_A - 1} \sigma_{B_i, j}^{z} \sigma_{A_i, j}^{y} \sigma_{B_{i-1}, j}^{x} \\
    & + \displaystyle\sum_{i = 1}^{N_A} \sigma_{A_i, j}^{x} \sigma_{B_{i-1}, j}^{z} \sigma_{A_{i}, j-1}^{y} \\
    & + \displaystyle\sum_{i = 1}^{N_A - 1} \sigma_{B_{i-1}, j+ 1}^{y} \sigma_{A_i, j}^{x} \sigma_{B_i, j}^{z} \\
    & \left. + \displaystyle\sum_{i = 1}^{N_A - 1} \sigma_{A_i, j}^{z} \sigma_{B_i, j}^{y} \sigma_{A_{i+1}, j}^{x} \right).
\end{aligned}
\end{equation}
It should be noted that in Majorana fermion representation the modes $b_{B_0, j}^{z}$ and $b_{A_{N_A}, j}^{z}$ are part of the extended Hilbert space $\tilde{\mathcal{H}}$, but do not couple to any other modes (see Fig.~\ref{fig:kitaev_zigzag}) and hence engender energetic zero modes. It can however be argued that the TRI breaking Term $\hat{H}_{K}$ is the effective low energy Hamiltonian expanded from~\cite{Kit}
\begin{equation}
    \hat{H}_{h} = \displaystyle\sum_{\vec{R}, \gamma = A, B} h_x \sigma_{\gamma, \vec{R}}^{x} + h_y \sigma_{\gamma, \vec{R}}^{y} + h_z \sigma_{\gamma, \vec{R}}^{z},
\end{equation}
with $K = h_x h_y h_z / J^2$. In this case the $b_{B_0, j}^{z}$ and $b_{A_{N_A}, j}^{z}$ modes couple linearly in $h_z$ to $c_{B_0, j}$ and $c_{A_{N_A}, j}$. 

Defining $a_x$ to be the $x$-component of $\vec{a}_1 - \vec{a}_2$, to diagonalize the Hamiltonian we define the Fourier transform
\begin{equation}
    c_{\gamma, j} = \frac{1}{\sqrt{N_x}} \displaystyle\sum_{k_x} \e^{-\ci k_x a_x j} c_{\gamma, k_x}
\end{equation}
(for the $b^{z}$ modes analogously) and we can write the Hamiltonian as
\begin{equation}\label{Eq:fourier_transformed_open_boundaries}
    \hat{\tilde{H}}_{u = 1} + \hat{\tilde{H}}_{K, u = 1, D = 1} = \displaystyle\sum_{k_x} \vec{c}^{\dagger}_{k_x} \tilde{h}(k_x) \vec{c}_{k_x},
\end{equation}
where $\vec{c}_{k_x} = (b^z_{B_0, k_x}, c_{B_0, k_x}, c_{A_1, k_x}, \dots)^{T}$ and
\begin{equation}\label{Eq:hamiltonian_matrix_open_boundaries}
\tilde{h}(k_x) = \begin{pmatrix}
    0 & \frac{\ci h_z}{2} &  & & & & \\
    - \frac{\ci h_z}{2} & \alpha_{k_x} & \gamma_{k_x}^{*} & \beta_{k_x}^{*} & & & \\
    & \gamma_{k_x} & -\alpha_{k_x} & \frac{\ci J_z}{2} & - \beta_{k_x}^{*} & & \\
    & \beta_{k_x} & -\frac{\ci J_z}{2} & \alpha_{k_x} & \gamma_{k_x}^{*} & \beta_{k_x}^{*} & \\
    & & -\beta_{k_x} & \gamma_{k_x} & - \alpha_{k_x} & \frac{\ci J_z}{2} & \ddots \\
    & & & \beta_{k_x} & \frac{-\ci J_z}{2} & \alpha_{k_x} & \ddots \\
    & & & & \ddots & \ddots & \ddots
\end{pmatrix},
\end{equation}
with
$\alpha_{k_x} = K \sin k_x a_x$, $\gamma_{k_x} = \ci (J_x + J_y \e^{- \ci k_x a_x})/2$ and $\beta_{k_x} = \ci K (1 - \e^{-\ci k_x a_x})/2$.

\subsection{Eigenmodes}\label{App:third_2_appendix}

From Eq.~\eqref{Eq:hamiltonian_matrix_open_boundaries} we can obtain the energy spectrum. How do we obtain the eigenmodes? It is possible to block diagonalize a Hamiltonian of the form of Eq.~\eqref{Eq:app_majorana_hamiltonian}, where $A$ is some $2N \times 2N$ matrix into $2 \times 2$ matrices with an orthogonal transformation $Q$, so that $Q \vec{c}$ again describes Majorana modes. Diagonalizing the $2 \times 2$ blocks with a unitary transformation results in a total of $N$ orthonormal ``complex'' fermions obtained from the ``real'' Majorana fermions. Since the Kitaev model is translation invariant, it is expedient to first perform a Fourier transformation leading to the set of modes $\cup_{k_x} \lbrace \bar{c}_{1, k_x}, \dots, \bar{c}_{2N_A + 2, k_x} \rbrace$ (where $\bar{c}_{j, k_x}$ is a more compact representation for the modes $\bar{c}_{2 i + j, k_x} := c_{A_j, k_x}, \bar{c}_{2j + 2, k_x} := c_{B_j, k_x}, \bar{c}_{1, k_x} := b^{z}_{B_0, k_x}$ and $\bar{c}_{2N_A + 2, k_x} := b^{z}_{A_{N_A}, k_x}$, in particular when summing over the modes). It is important to note though that the Fourier-transformed modes are neither Majorana fermions, nor proper complex fermions due to the commutator relation
\begin{equation}
    \lbrace \bar{c}_{i, k_x}^{\dagger}, \bar{c}_{j, k_x'} \rbrace =  \lbrace \bar{c}_{i, -k_x}, \bar{c}_{j, k_x'} \rbrace = 2 \delta_{ij} \delta_{k_x, k_x'}.
\end{equation}
For this reason, it is in general not possible to transform the $2N_A+2$ modes $\lbrace \bar{c}_{j, k_x} \rbrace$ into $N_A+1$ complex fermionic modes, as for $k_x \neq -k_x (\text{mod } 2 \pi)$ the adjoints of $\lbrace \bar{c}_{j, k_x} \rbrace$ cannot be written as a superposition of the $\bar{c}_{j, k_x}$ operators themselves. Instead, we need to combine modes at $k_x$ and $-k_x$.

To obtain the eigenmodes, first we need to diagonalize the Hamiltonian matrix in Eq.~\eqref{Eq:hamiltonian_matrix_open_boundaries} in a way to take into account that we want to construct eigenmodes using the $\bar{c}_{j, k_x}$ operators at $k_x$ and $-k_x$. Let $U_{D}^{k_x}$ be a unitary matrix that diagonalizes $\tilde{h}(k_x)$ for a given $k_x$. There are many unitary matrices that are similar to each other that diagonalize a given hermitian matrix. First, we require that
\begin{equation}\label{Eq:diagonalization_convention}
\begin{aligned}[b]
    U_{D}^{k_x} \tilde{h}(k_x) \left( U_{D}^{k_x} \right)^{\dagger} = \begin{pmatrix}
        \epsilon_{1, k_x} & & & & \\
        & \epsilon_{2, k_x} & & & \\
        & & \ddots & & \\
        & & & -\epsilon_{2, k_x} & \\
        & & & & - \epsilon_{1, k_x}
    \end{pmatrix} \\
    = \epsilon(k_x),
\end{aligned}
\end{equation}
so that the eigenvalues $\epsilon_{j, k_x}$ are sorted by size with $\epsilon_{1, k_x} \geq \epsilon_{2, k_x} \geq \dots \geq -\epsilon_{2, k_x} \geq - \epsilon_{1, k_x}$. We can then relate $U_{D}^{k_x}$ and $U_{D}^{-k_x}$ by noting two symmetries. According to Table~\ref{tab:space_groups} the Hamiltonian is always inversion symmetric. Inversion acts on the spins as
\begin{equation}
    \begin{aligned}[b]
        I: & (A_i, j) \mapsto (B_{N_A - i}, -j) \\
        & (B_i, j) \mapsto (A_{N_A - i, -j}) \\
        & \sigma_{\gamma, j}^{\alpha} \mapsto \sigma_{\bar{\gamma}, -j}^{\alpha},
    \end{aligned}
\end{equation}
(with $\bar{\gamma}$ denoting the sublattice obtained by inversion) because spin operators transform like pseudovectors. If however the spin operators acted on the Majorana operators like
\begin{equation}
    \begin{aligned}[b]
    I: & c_{\gamma, j} \mapsto c_{\bar{\gamma}, -j} \\
    & b_{\gamma, j}^{\alpha} \mapsto b_{\bar{\gamma}, -j}^{\alpha},
    \end{aligned}
\end{equation}
then the link operators $\hat{u}_{\vec{R}, \vec{R}'}$ would flip their sign under inversion. This issue can be avoided by requiring that the Majorana operators transform projectively under inversion~\cite{Zha}:
\begin{equation}
    \begin{aligned}[b]
        I: & c_{A_i, j} \mapsto - c_{B_{N_A} - i, -j} \\
        & b_{A_i, j}^{\alpha} \mapsto - b_{B_{N_A}- i, -j}^{\alpha} \\
        & c_{B_i, j} \mapsto c_{A_{N_A} - i, -j} \\
        & b_{B_i, j}^{\alpha} \mapsto b_{A_{N_A} - i, -j}^{\alpha}.
    \end{aligned}
\end{equation}
In matrix representation, acting on the $\vec{c}_{k_x}$ modes like $I_M: \vec{c}_{k_x} \mapsto I_M \vec{c}_{k_x}$ we can write the inversion operator as
\begin{equation}
    (I_M)_{ij} = \eta_{i} (-1)^{i} \delta_{j, (2 N_A + 2) + 1 - i},
\end{equation}
with $\eta_{i} = -1$ if $i = 1, 2N_A + 2$ and $=1$ in all other cases. One readily finds $I_M \tilde{h}(k_x) I_M^{\dagger}  = \tilde{h}(-k_x)$, so the eigenspectra at $k_x$ and $-k_x$ are identical, as both operators are related to each other by a unitary symmetry.

The second symmetry is particle-hole symmetry. Since Majorana operators are their own antiparticles, a particle-hole transformation leaves Majoranas invariant, however it acts as complex conjugation on complex numbers as it is an antiunitary symmetry. From Eq.~\eqref{Eq:app_majorana_hamiltonian} we can immediately tell that particle-hole conjugation anticommutes with the Hamiltonian, and comparing with Eq.~\eqref{Eq:fourier_transformed_open_boundaries} we identify
\begin{equation}
    \tilde{h}(k_x) = - \tilde{h}(-k_x)^{*}.
\end{equation}

It is now possible to relate $U_{D}^{k_x}$ with $U_{D}^{-k_x}$. Due to the fact that the spectra of $\tilde{h}(k_x)$ and $\tilde{h}(-k_x)$ are identical, we may write
\begin{equation}
\begin{aligned}[b]
    \epsilon(k_x) = & U_{D}^{k_x} \tilde{h}(k_x) \left( U_{D}^{k_x} \right)^{\dagger} = U_{D}^{-k_x} \tilde{h}(-k_x) \left( U_{D}^{-k_x} \right)^{\dagger} \\
    = & - \left( U_{D}^{-k_x} \right)^{*} \tilde{h}(k_x) \left( \left( U_{D}^{-k_x} \right)^{\dagger} \right)^{*} \\
    = & \Sigma_x \left( U_{D}^{-k_x} \right)^{*} \tilde{h}(k_x) \left( \left( U_{D}^{-k_x} \right)^{\dagger} \right)^{*} \Sigma_x,
 \end{aligned}
\end{equation}
where $(\Sigma_x)_{ij} = \delta_{i, (2 N_{A} + 2) + 1 - j}$; therefore we identify
\begin{equation}\label{Eq:pos_neg_k_x_condition}
    U_{D}^{-k_x} = \Sigma_{x} \left( U_{D}^{k_x} \right)^{*}.
\end{equation}
By defining complex modes $\bar{a}_{i, k_x}$ via
\begin{equation}
    \bar{a}_{i, k_x} = \frac{1}{\sqrt{2}} \displaystyle\sum_{j = 1}^{2N_A + 2} \left( U_{D}^{k_x} \right)_{ij} \bar{c}_{j, k_x}
\end{equation}
one quickly finds $\lbrace \bar{a}_{i, k_x}^{\dagger}, \bar{a}_{j, k_x'} \rbrace = \delta_{k_x, k_x'} \delta_{i,j}$ and
\begin{equation}
    \begin{aligned}[b]
    \lbrace \bar{a}_{i, k_x}, \bar{a}_{j, k_x'} \rbrace = & \frac{1}{2} \left( U_{D}^{k_x} \right)_{i \alpha} \left( U_{D}^{k_x'} \right)_{j \beta} \lbrace \bar{c}_{\alpha, k_x}, \bar{c}_{\beta, k_x'} \rbrace \\
    = & \delta_{k_x, -k_x'} \left( U_{D}^{k_x} \right)_{i \alpha} \left( U_{D}^{-k_x} \right)_{j \alpha} \\
    = & \delta_{k_x, -k_x'} \left( U_{D}^{k_x} \right)_{i \alpha} \left( \left( U_{D}^{k_x} \right)^{\dagger} \Sigma_{x} \right)_{\alpha j} \\
    = & \delta_{k_x, -k_x'} \delta_{i, (2 N_A + 2) + 1 - j},
    \end{aligned}
\end{equation}
so we identify
\begin{equation}
    \bar{a}_{i, k_x} = \bar{a}_{(2 N_A + 2) + 1 - i, -k_x}^{\dagger}.
\end{equation}
This we write as
\begin{equation}
    \begin{pmatrix}
        a_{1, k_x} \\
        a_{2, k_x} \\
        \vdots \\
        a_{2, -k_x}^{\dagger} \\
        a_{1, -k_x}^{\dagger}
    \end{pmatrix} = \frac{1}{\sqrt{2}} U_{D}^{k_x} \vec{c}_{k_x},
\end{equation}
where we have defined $a_{i, k_x} = \bar{a}_{i, k_x}$ for $1 \leq i \leq N_A + 1$. Note that in this form Eq.~\eqref{Eq:fourier_transformed_open_boundaries} becomes
\begin{equation}
    \displaystyle\sum_{k_x} \vec{c}_{k_x}^{\dagger} \tilde{h}(k_x) \vec{c}_{k_x} =4 \displaystyle\sum_{k_x} \displaystyle\sum_{1 \leq i \leq N_A + 1} \epsilon_{i, k_x} \left( a_{i, k_x}^{\dagger} a_{i, k_x} - \frac{1}{2} \right),
\end{equation}
so that the ground state is still the vacuum state.

In the case $k_x = \bar{k}_x$ (where $\bar{k}_x = -\bar{k}_x$ mod $2\pi$, i.e. $\bar{k}_x = 0, \pi$) we get a slightly different condition: Eq.~\eqref{Eq:pos_neg_k_x_condition} requires
\begin{equation}\label{Eq:bar_k_x_condition}
    \left( U_{D}^{\bar{k}_x} \right)_{ij} = \left( U_{D}^{\bar{k}_x} \right)^{*}_{(2 N_A + 2) + 1 - i, j},
\end{equation}
which can be realized numerically by computing some unitary matrix that diagonalizes $\tilde{h}(\bar{k}_x)$ with the same convention as in Eq.~\eqref{Eq:diagonalization_convention} and only keeping the first $N_A + 1$ rows of $U_{D}^{\bar{k}_x}$, while constructing the remaining ones according to Eq.~\eqref{Eq:bar_k_x_condition}.

\subsection{Expansion of the Hamiltonian}\label{App:third_3_appendix}

For a stripe geometry we define, based on the different ``orbitals'' in Eq.~\eqref{Eq:orbitals}

\begin{equation}\label{Eq:orbitals_open_boundaries}
    \begin{aligned}[b]
        & \hat{O}^{x, i}_{j} = \sigma_{2 i, j}^{x} \sigma_{2 i - 1, j}^{x}, \qquad && i \in \lbrace 1, \dots, N_A \rbrace \\
        & \hat{O}^{y, i}_{j} = \sigma_{2 i, j}^{y} \sigma_{2 i - 1, j + 1}^{y}, \qquad && i \in \lbrace 1, \dots, N_A \rbrace \\
        & \hat{O}^{z, i}_{j} = \sigma_{2 i, j}^{z} \sigma_{2 i + 1, j}^{z}, \qquad && i \in \lbrace 1, \dots, N_A - 1 \rbrace \\  
        & \hat{O}^{A_1, i}_{j} = \sigma_{2 i, j}^{z} \sigma_{2 i + 1, j}^{y} \sigma_{2 i + 2, j}^{x}, \qquad && i \in \lbrace 1, \dots, N_A - 1 \rbrace \\
        & \hat{O}^{A_2, i}_{j} = \sigma_{2 i +2, j-1}^{y} \sigma_{2 i + 1, j}^{x} \sigma_{2 i, j}^{z}, \qquad && i \in \lbrace 1, \dots, N_A - 1 \rbrace \\
        & \hat{O}^{A_3, i}_{j} = \sigma_{2 i, j}^{x} \sigma_{2 i - 1, j}^{z} \sigma_{2 i, j-1}^{y}, \qquad && i \in \lbrace 1, \dots, N_A \rbrace \\
        & \hat{O}^{B_1, i}_{j} = \sigma_{2 i+1, j}^{z} \sigma_{2 i, j}^{y} \sigma_{2 i - 1, j}^{x}, \qquad && i \in \lbrace 1, \dots, N_A - 1 \rbrace \\
        & \hat{O}^{B_2, i}_{j} = \sigma_{2 i-1, j+1}^{y} \sigma_{2 i, j}^{x} \sigma_{2 i + 1, j}^{z}, \qquad && i \in \lbrace 1, \dots, N_A - 1 \rbrace \\
        & \hat{O}^{B_3, i}_{j} = \sigma_{2 i - 1, j}^{x} \sigma_{2 i, j}^{z} \sigma_{2 i - 1, j+1}^{y}, \qquad && i \in \lbrace 1, \dots, N_A \rbrace
    \end{aligned}
\end{equation}
where we use the convention $A_i = 2i$ and $B_i = 2i+1$. In addition, the operators
\begin{equation}\label{Eq:sigma_z_spin_orbitals}
    \begin{aligned}[b]
        \hat{O}_{j}^{b^{z}_{B_0}} = & \sigma^{z}_{1, j} \\
        \hat{O}_{j}^{b^{z}_{A_N}} = & \sigma^{z}_{2 N_A, j}        
    \end{aligned}
\end{equation}
will only engender itinerant Majorana modes, too, as the modes $b_{B_0}^{z}$ and $b_{A_{N_A}}^{z}$ have no other $b^{z}$ modes to form a link operator and hence do not change the flux configuration of a state. Similarly to Eq.~\eqref{Eq:orbitals_on_ground_state} we can express the action of the operators in Eq.~\eqref{Eq:orbitals_open_boundaries} on the vacuum state as
\begin{equation}
\begin{aligned}[b]
    \hat{O}^{\mu}_{q_x} \to \frac{1}{\sqrt{N_x}} \displaystyle\sum_{k_x} \displaystyle\sum_{\alpha, \beta = 1}^{N_A+1} & \left( \Omega^{\mu, \alpha, \beta}_{k_x, q_x} \left( a_{k_x}^{\alpha} \right)^{\dagger} \left( a_{-k_x-q_x}^{\beta} \right)^{\dagger} \right. \\
    & \left. + \bar{\Omega}^{\mu, \alpha, \beta}_{k_x, q_x} \delta_{q_x, 0} \right).
\end{aligned}
\end{equation}
One identifies

\begin{equation*}
    \begin{aligned}
        \Omega^{(x,i), \alpha, \beta}_{k_x, q_x} = & -2 \ci \left( \left( U_{D}^{k_x} \right)^{\dagger} \right)^{*}_{2i+1, \alpha} \\
        & \cdot \left( \left( U_{D}^{k_x + q_x} \right)^{\dagger} \right)_{2i, (2 N_A + 2) + 1 - \beta} \\
        \Omega^{(y,i), \alpha, \beta}_{k_x, q_x} = & -2 \ci \e^{-\ci (k_x + q_x) a_x} \left( \left( U_{D}^{k_x} \right)^{\dagger} \right)^{*}_{2i+1, \alpha} \\ 
        & \cdot \left( \left( U_{D}^{k_x + q_x} \right)^{\dagger} \right)_{2i, (2 N_A + 2) + 1 - \beta} \\
        \Omega^{(z,i), \alpha, \beta}_{k_x, q_x} = & -2 \ci \left( \left( U_{D}^{k_x} \right)^{\dagger} \right)^{*}_{2i+1, \alpha} \\ 
        & \cdot \left( \left( U_{D}^{k_x + q_x} \right)^{\dagger} \right)_{2i + 2, (2 N_A + 2) + 1 - \beta} \\
    \end{aligned}
\end{equation*}
\begin{equation}
    \begin{aligned}[b]
        \Omega^{(A_1,i), \alpha, \beta}_{k_x, q_x} = & -2 \ci \left( \left( U_{D}^{k_x} \right)^{\dagger} \right)^{*}_{2i+1, \alpha} \\ 
        & \cdot \left( \left( U_{D}^{k_x + q_x} \right)^{\dagger} \right)_{2i +3, (2 N_A + 2) + 1 - \beta} \\
        \Omega^{(A_2,i), \alpha, \beta}_{k_x, q_x} = & -2 \e^{-\ci k_x a_x} \ci \left( \left( U_{D}^{k_x} \right)^{\dagger} \right)^{*}_{2i+3, \alpha} \\ 
        & \cdot \left( \left( U_{D}^{k_x + q_x} \right)^{\dagger} \right)_{2i+1, (2 N_A + 2) + 1 - \beta} \\
        \Omega^{(A_3,i), \alpha, \beta}_{k_x, q_x} = & -2 \e^{\ci (k_x + q_x) a_x} \ci \left( \left( U_{D}^{k_x} \right)^{\dagger} \right)^{*}_{2i+1, \alpha} \\ 
        & \cdot \left( \left( U_{D}^{k_x + q_x} \right)^{\dagger} \right)_{2i+1, (2 N_A + 2) + 1 - \beta} \\
        \Omega^{(B_1,i), \alpha, \beta}_{k_x, q_x} = & -2 \ci \left( \left( U_{D}^{k_x} \right)^{\dagger} \right)^{*}_{2i+2, \alpha} \\ 
        & \cdot \left( \left( U_{D}^{k_x + q_x} \right)^{\dagger} \right)_{2i, (2 N_A + 2) + 1 - \beta} \\
        \Omega^{(B_2,i), \alpha, \beta}_{k_x, q_x} = & -2 \e^{\ci k_x a_x} \ci \left( \left( U_{D}^{k_x} \right)^{\dagger} \right)^{*}_{2i, \alpha} \\ 
        & \cdot \left( \left( U_{D}^{k_x + q_x} \right)^{\dagger} \right)_{2i+2, (2 N_A + 2) + 1 - \beta} \\
        \Omega^{(B_3,i), \alpha, \beta}_{k_x, q_x} = & -2 \ci \e^{-\ci (k_x + q_x) a_x} \left( \left( U_{D}^{k_x} \right)^{\dagger} \right)^{*}_{2i, \alpha} \\ 
        & \cdot \left( \left( U_{D}^{k_x + q_x} \right)^{\dagger} \right)_{2i, (2 N_A + 2) + 1 - \beta} \\
        \Omega^{b_{B_0}^{z}, \alpha, \beta}_{k_x, q_x} = & 2 \ci \left( \left( U_{D}^{k_x} \right)^{\dagger} \right)^{*}_{1, \alpha} \\ 
        & \cdot \left( \left( U_{D}^{k_x + q_x} \right)^{\dagger} \right)_{2, (2 N_A + 2) + 1 - \beta} \\
        \Omega^{b_{A_N}^{z}, \alpha, \beta}_{k_x, q_x} = & 2 \ci \left( \left( U_{D}^{k_x} \right)^{\dagger} \right)^{*}_{2N_A+2, \alpha} \\ 
        & \cdot \left( \left( U_{D}^{k_x + q_x} \right)^{\dagger} \right)_{2N_A + 1, (2 N_A + 2) + 1 - \beta}
    \end{aligned}
\end{equation}
For the expanded Hamiltonian we identify
\begin{equation}\label{Eq:hamiltonian_expansion_open_boundaries}
    \begin{aligned}[b]
        H^{\mu \nu}_{q_x} = & \displaystyle\sum_{\alpha, \beta} \left( S^{-\frac{1}{2}}_{q_x} \right)_{\mu \alpha} \left( S^{-\frac{1}{2}}_{q_x} \right)_{\beta \nu} \\
        & \cdot \frac{1}{N_x} \displaystyle\sum_{k_x} \displaystyle\sum_{\gamma, \delta = 1}^{N_A + 1} (4\epsilon_{\gamma, k_x} + 4 \epsilon_{\delta, k_x + q_x}) \\
        & \cdot \left( \left( \Omega^{\alpha, \gamma, \delta}_{k_x, q_x} \right)^{*} \Omega^{\beta, \gamma, \delta}_{k_x, q_x} - \left( \Omega^{\alpha, \gamma, \delta}_{k_x, q_x} \right)^{*} \Omega^{\beta, \delta, \gamma}_{-k_x-q_x, q_x} \right).
    \end{aligned}
\end{equation}

\section{Hamiltonian expansion for nine spin orbitals}\label{App:fourth_appendix}

\begin{figure}[ht]
 	\centering
 	\includegraphics[width=1.0\linewidth]{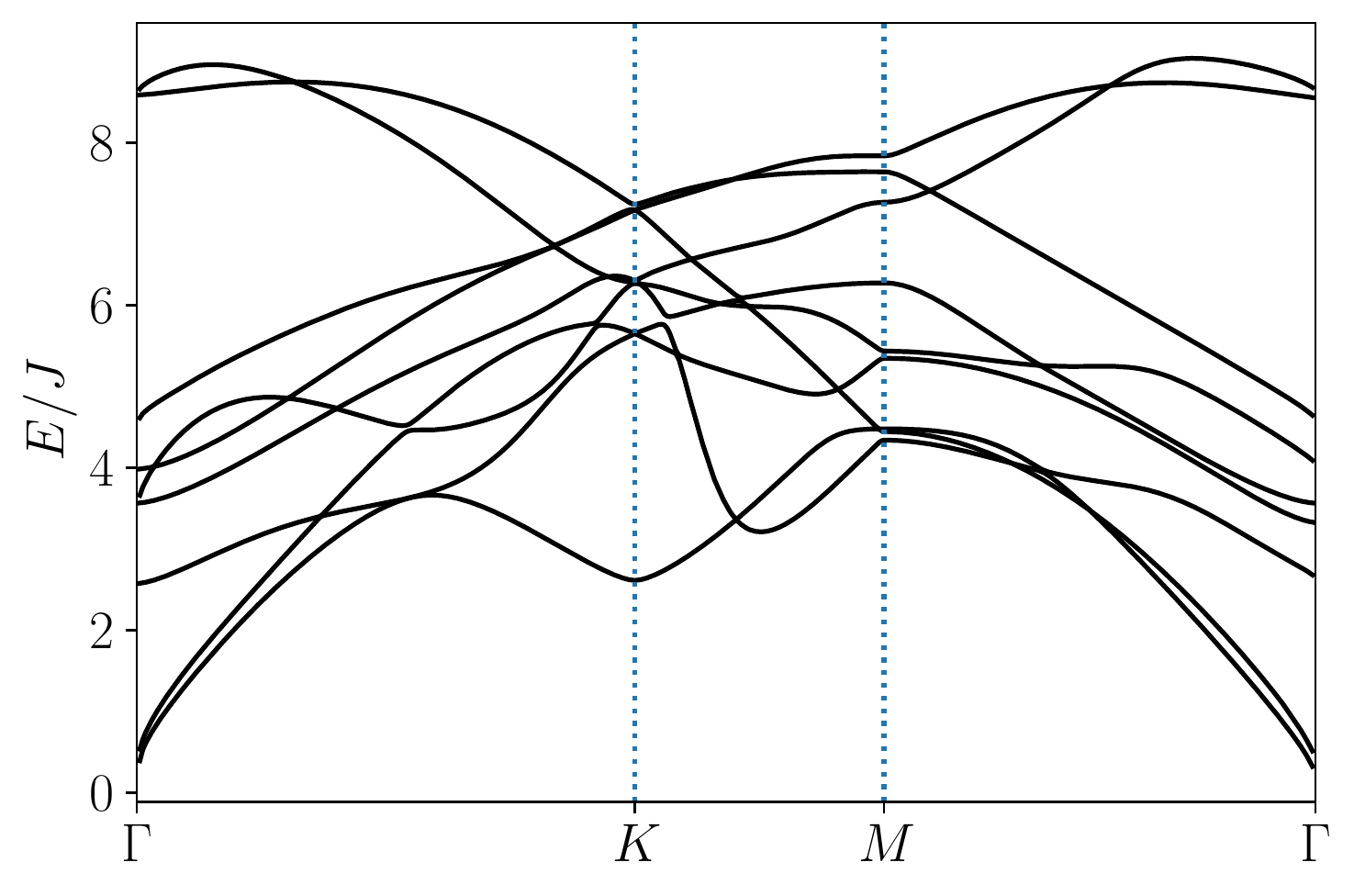}
 	\caption{Eigenvalue spectrum of $H_{\vec{q}}$ (infinite system) for isotropic exchange coupling $J = 1$ and $K = 0$ for all spin orbitals in Eq.~\eqref{Eq:orbitals}.}
 	\label{fig:H_exp_9_orb_K_0.0_J_1}
 \end{figure}

\begin{figure}[ht]
 	\centering
 	\includegraphics[width=1.0\linewidth]{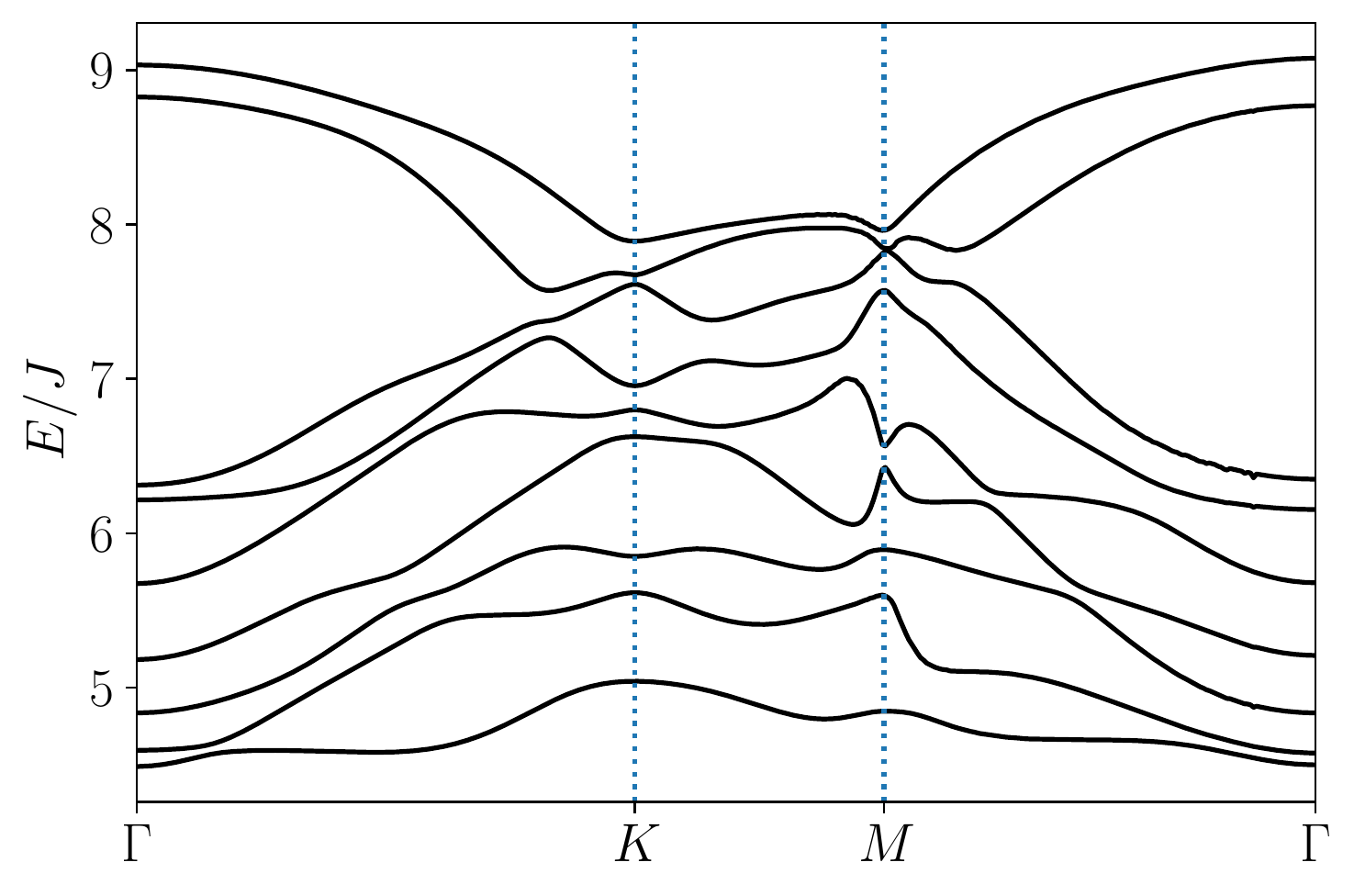}
 	\caption{Eigenvalue spectrum of $H_{\vec{q}}$ (infinite system) for isotropic exchange coupling $J = 1$ and $K = 0.2$ for all spin orbitals in Eq.~\eqref{Eq:orbitals}.}
 	\label{fig:H_exp_9_orb_K_0.2_J_1_modified}
 \end{figure}

\begin{figure}[ht]
 	\centering
 	\includegraphics[width=1.0\linewidth]{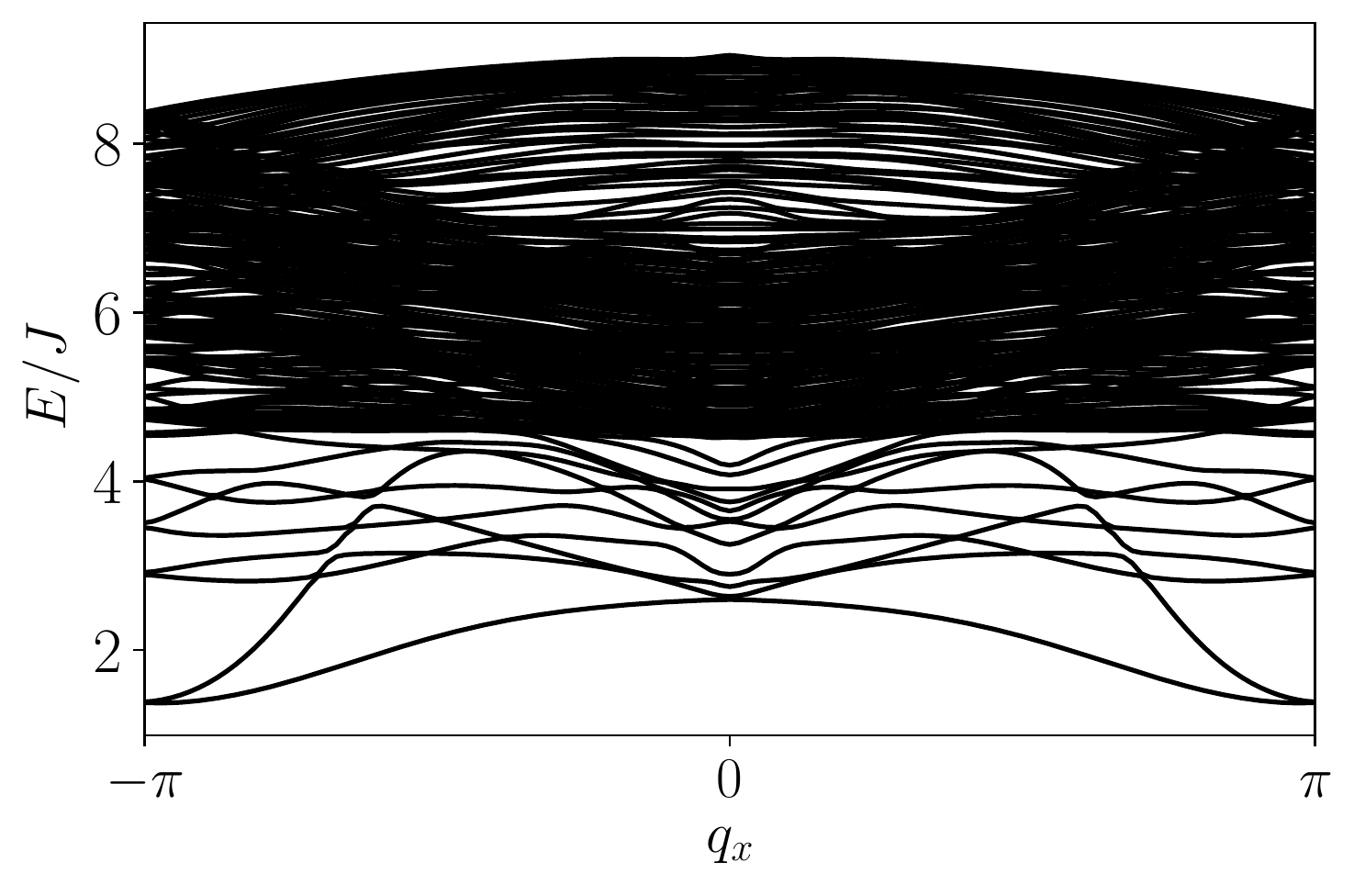}
 	\caption{Eigenvalue spectrum of $H_{q_x}$ (see Eq.~\eqref{Eq:hamiltonian_expansion_open_boundaries}) for isotropic exchange coupling $J = 1$ and $K = 0.2$. All of the spin orbitals in Eq.~\eqref{Eq:orbitals_open_boundaries} are considered.}
 	\label{fig:H_exp_edge_3_orb_K_0.2_J_1}
 \end{figure}

\begin{figure}[ht]
 	\centering
 	\includegraphics[width=1.0\linewidth]{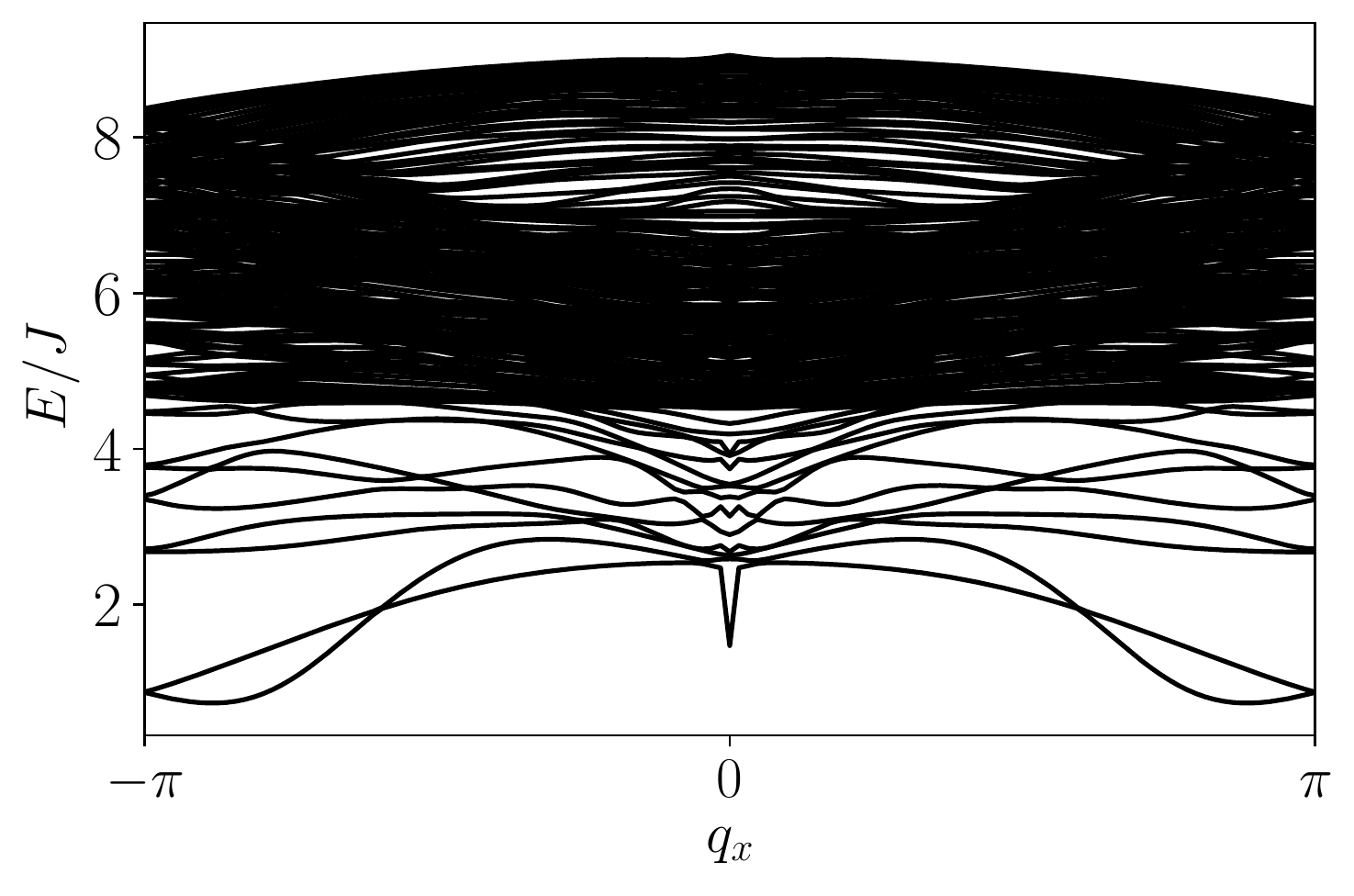}
 	\caption{Eigenvalue spectrum of $H_{q_x}$ (see Eq.~\eqref{Eq:hamiltonian_expansion_open_boundaries}) for isotropic exchange coupling $J = 1$ and $K = 0.2$. All of the spin orbitals in Eq.~\eqref{Eq:orbitals_open_boundaries} and Eq.~\eqref{Eq:sigma_z_spin_orbitals} are considered.}
 	\label{fig:H_exp_edge_3_with_b_z_orb_K_0.2_J_1}
 \end{figure}

Plots for the expansion of the Hamiltonian with nine spin orbitals are shown in Figs.~\ref{fig:H_exp_9_orb_K_0.0_J_1}, \ref{fig:H_exp_9_orb_K_0.2_J_1_modified}, \ref{fig:H_exp_edge_3_orb_K_0.2_J_1} and \ref{fig:H_exp_edge_3_with_b_z_orb_K_0.2_J_1}.

\section{Spin operator ``orbitals''}\label{App:fifth_appendix}

\subsection{Action of ``orbitals'' on the ground state}\label{App:fourth_1_appendix}

The action of the ``orbitals'' defined in Eq.~\eqref{Eq:orbitals} on the ground state, setting both the link operators $\hat{u}^{\alpha}_{\vec{R}, \vec{R}'}$ and gauge operators $\hat{D}_{\gamma, \vec{R}}$ equal to one is given by Eq.~\eqref{Eq:orbitals_on_ground_state}, where
\begin{equation}\label{Eq:omegas}
    \begin{aligned}[b]
        \Omega^{x}_{\vec{k}, \vec{q}} = & - 2 \ci \e^{\ci \vec{k} \vec{a}_1 + \ci \vec{q} \vec{a}_1} \e^{- \ci \phi_{\vec{k}}} \sin \varphi_{\vec{k}} \sin \varphi_{\vec{k} + \vec{q}}, \\
        \Omega^{y}_{\vec{k}, \vec{q}} = & - 2 \ci \e^{\ci \vec{k} \vec{a}_2 + \ci \vec{q} \vec{a}_2} \e^{- \ci \phi_{\vec{k}}} \sin \varphi_{\vec{k}} \sin \varphi_{\vec{k} + \vec{q}}, \\
        \Omega^{z}_{\vec{k}, \vec{q}} = & - 2 \ci \e^{- \ci \phi_{\vec{k}}} \sin \varphi_{\vec{k}} \sin \varphi_{\vec{k} + \vec{q}}, \\
        \Omega^{A_1}_{\vec{k}, \vec{q}} = & 2 \ci \e^{-\ci \vec{k} \vec{a}_1 - \ci \vec{q} \vec{a}_1} \e^{- \ci \phi_{\vec{k}}} \e^{ \ci \phi_{\vec{k} + \vec{q}}} \sin \varphi_{\vec{k}} \cos \varphi_{\vec{k} + \vec{q}}, \\
        \Omega^{A_2}_{\vec{k}, \vec{q}} = & 2 \ci \e^{\ci \vec{k} \vec{a}_2} \e^{- \ci \phi_{\vec{k}}} \e^{ \ci \phi_{\vec{k} + \vec{q}}} \sin \varphi_{\vec{k}} \cos \varphi_{\vec{k} + \vec{q}}, \\
        \Omega^{A_3}_{\vec{k}, \vec{q}} = & 2 \ci \e^{\ci \vec{k} (\vec{a}_1 - \vec{a}_2) - \ci \vec{q} \vec{a}_2} \e^{- \ci \phi_{\vec{k}}} \e^{ \ci \phi_{\vec{k} + \vec{q}}} \sin \varphi_{\vec{k}} \cos \varphi_{\vec{k} + \vec{q}}, \\
        \Omega^{B_1}_{\vec{k}, \vec{q}} = & - 2 \ci \e^{\ci \vec{k} \vec{a}_1 + \ci \vec{q} \vec{a}_1} \cos \varphi_{\vec{k}} \sin \varphi_{\vec{k} + \vec{q}}, \\
        \Omega^{B_2}_{\vec{k}, \vec{q}} = & - 2 \ci \e^{-\ci \vec{k} \vec{a}_2} \cos \varphi_{\vec{k}} \sin \varphi_{\vec{k} + \vec{q}}, \\
        \Omega^{B_3}_{\vec{k}, \vec{q}} = & - 2 \ci \e^{-\ci \vec{k} (\vec{a}_1 - \vec{a}_2 ) + \ci \vec{q} \vec{a}_2} \cos \varphi_{\vec{k}} \sin \varphi_{\vec{k} + \vec{q}}
    \end{aligned}
\end{equation}

\subsection{Transformation of ``orbitals'' under unitary symmetries}\label{App:fourth_2_appendix}

Below, we list the way the ``orbitals'' defined in Eq.~\eqref{Eq:orbitals} transform under the action of unitary symmetries at the three high symmetry points $\Gamma = \vec{0}$, $K = (\vec{b}_1 - \vec{b}_2)/3$ and $M = \vec{b}_1/2$ with the reciprocal lattice vectors $\vec{b}_1 = (2\pi, 2\pi/\sqrt{3})$ and $\vec{b}_2 = (-2\pi, 2\pi/\sqrt{3})$. We choose the basis $(xx, yy, zz, A_1, A_2, A_3, B_1, B_2. B_3)$ for the matrix representation $T_{\vec{k}}$ of the symmetry elements at wave vector $\vec{k}$, so that
\begin{equation}\label{Eq:action_of_symmetry_on_orbital}
    \hat{g} \left( \hat{O}^{\mu}_{\vec{q}} \right)^{\dagger} \hat{g}^{\dagger} = T_{\vec{q}}^{\mu \nu}(g) \left( \hat{O}^{\nu}_{\vec{q}} \right)^{\dagger},
\end{equation}
where $\hat{g}$ is an element of the little cogroup $\tilde{G}_{\vec{q}}$. Eq.~\eqref{Eq:action_of_symmetry_on_orbital} holds analogously for $\hat{\bar{O}}^{\mu}_{\vec{q}}$ and it is straightforward to check that $T_{\vec{q}}(g)$ commutes with $H_{\vec{q}}$. For simplicity we define $\omega = \e^{2\pi \ci/3}$ and $\omega^{*} = \e^{-2 \pi \ci / 3}$ and matrices are divided into $3 \times 3$ blocks to guide the eye.

\begin{equation}
    T_{\Gamma}(E) = \left( \begin{array}{ccc|ccc|ccc}
 1& & & & & & & & \\
 & 1& & & & & & & \\
 & & 1& & & & & & \\
 \hline
& & & 1& & & & & \\
& & & & 1& & & & \\
& & & & & 1& & &\\
\hline
& & & & & &1 & & \\
& & & & & & &1 & \\
& & & & & & & &1 \\
\end{array} \right)
\end{equation}

\begin{equation}
    T_{\Gamma}(C_3^{+}) = \left( \begin{array}{ccc|ccc|ccc}
 & 1& & & & & & & \\
 & & 1& & & & & & \\
 1& & & & & & & & \\
 \hline
& & & & & 1& & & \\
& & & 1& & & & & \\
& & & & 1& & & &\\
\hline
& & & & & & & & 1\\
& & & & & & 1& & \\
& & & & & & & 1& \\
\end{array} \right)
\end{equation}

\begin{equation}
    T_{\Gamma}(C_{3}^{-}) = \left( \begin{array}{ccc|ccc|ccc}
 & & 1& & & & & & \\
 1& & & & & & & & \\
 & 1& & & & & & & \\
 \hline
& & & &1 & & & & \\
& & & & &1 & & & \\
& & &1& & & & &\\
\hline
& & & & & & & 1& \\
& & & & & & & &1 \\
& & & & & & 1& & \\
\end{array} \right)
\end{equation}

\begin{equation}
    T_{\Gamma}(C_{21}') = \left( \begin{array}{ccc|ccc|ccc}
 & 1& & & & & & & \\
 1& & & & & & & & \\
 & & 1& & & & & & \\
 \hline
& & & &-1 & & & & \\
& & &-1 & & & & & \\
& & & & &-1 & & &\\
\hline
& & & & & & &-1 & \\
& & & & & &-1 & & \\
& & & & & & & &-1 \\
\end{array} \right)
\end{equation}

\begin{equation}
    T_{\Gamma}(C_{22}') = \left( \begin{array}{ccc|ccc|ccc}
 & & 1& & & & & & \\
 & 1& & & & & & & \\
1 & & & & & & & & \\
 \hline
& & &-1 & & & & & \\
& & & & &-1 & & & \\
& & & &-1 & & & &\\
\hline
& & & & & & -1& & \\
& & & & & & & &-1 \\
& & & & & & & -1& \\
\end{array} \right)
\end{equation}

\begin{equation}
    T_{\Gamma}(C_{23}') = \left( \begin{array}{ccc|ccc|ccc}
 1& & & & & & & & \\
 & & 1& & & & & & \\
 & 1& & & & & & & \\
 \hline
& & & & & -1& & & \\
& & & & -1& & & & \\
& & &-1 & & & & &\\
\hline
& & & & & & & & -1\\
& & & & & & & -1& \\
& & & & & & -1& & \\
\end{array} \right)
\end{equation}

\begin{equation}
    T_{\Gamma}(I) = \left( \begin{array}{ccc|ccc|ccc}
1 & & & & & & & & \\
 & 1& & & & & & & \\
 & & 1& & & & & & \\
 \hline
& & & & & & 1& & \\
& & & & & & & 1& \\
& & & & & & & &1\\
\hline
& & &1 & & & & & \\
& & & & 1& & & & \\
& & & & & 1& & & \\
\end{array} \right)
\end{equation}

\begin{equation}
    T_{\Gamma}(S_{6}^{+}) = \left( \begin{array}{ccc|ccc|ccc}
 & &1 & & & & & & \\
1 & & & & & & & & \\
 & 1& & & & & & & \\
 \hline
& & & & & & & 1& \\
& & & & & & & &1 \\
& & & & & & 1& &\\
\hline
& & & & 1& & & & \\
& & & & & 1& & & \\
& & & 1& & & & & \\
\end{array} \right)
\end{equation}

\begin{equation}
    T_{\Gamma}(S_{6}^{-}) = \left( \begin{array}{ccc|ccc|ccc}
 & 1& & & & & & & \\
 & &1 & & & & & & \\
 1& & & & & & & & \\
 \hline
& & & & & & & & 1\\
& & & & & & 1& & \\
& & & & & & & 1&\\
\hline
& & & & &1 & & & \\
& & &1 & & & & & \\
& & & &1 & & & & \\
\end{array} \right)
\end{equation}

\begin{equation}
    T_{\Gamma}(\sigma_{v1}) = \left( \begin{array}{ccc|ccc|ccc}
 &1 & & & & & & & \\
 1& & & & & & & & \\
 & & 1& & & & & & \\
 \hline
& & & & & & & -1& \\
& & & & & & -1& & \\
& & & & & & & &-1\\
\hline
& & & &-1 & & & & \\
& & & -1& & & & & \\
& & & & & -1& & & \\
\end{array} \right)
\end{equation}

\begin{equation}
    T_{\Gamma}(\sigma_{v2}) = \left( \begin{array}{ccc|ccc|ccc}
 1& & & & & & & & \\
 & & 1& & & & & & \\
 & 1& & & & & & & \\
 \hline
& & & & & & & & -1\\
& & & & & & & -1& \\
& & & & & & -1& &\\
\hline
& & & & &-1 & & & \\
& & & & -1& & & & \\
& & & -1& & & & & \\
\end{array} \right)
\end{equation}

\begin{equation}
    T_{\Gamma}(\sigma_{v3}) = \left( \begin{array}{ccc|ccc|ccc}
 & & 1& & & & & & \\
 & 1& & & & & & & \\
 1& & & & & & & & \\
 \hline
& & & & & &-1 & & \\
& & & & & & & &-1 \\
& & & & & & & -1&\\
\hline
& & & -1& & & & & \\
& & & & & -1& & & \\
& & & & -1& & & & \\
\end{array} \right)
\end{equation}

\begin{equation}
    T_{K}(E) = \left( \begin{array}{ccc|ccc|ccc}
1 & & & & & & & & \\
 & 1& & & & & & & \\
 & & 1& & & & & & \\
 \hline
& & & 1& & & & & \\
& & & &1 & & & & \\
& & & & &1 & & &\\
\hline
& & & & & &1 & & \\
& & & & & & &1 & \\
& & & & & & & &1 \\
\end{array} \right)
\end{equation}

\begin{equation}
    T_{K}(C_{3}^{+}) = \left( \begin{array}{ccc|ccc|ccc}
 & \omega^{*}& & & & & & & \\
 & & \omega^{*}& & & & & & \\
 \omega^{*}& & & & & & & & \\
 \hline
& & & & &\omega & & & \\
& & &\omega & & & & & \\
& & & &\omega & & & &\\
\hline
& & & & & & & &\omega^{*} \\
& & & & & & \omega^{*}& & \\
& & & & & & &\omega^{*} & \\
\end{array} \right)
\end{equation}

\begin{equation}
    T_{K}(C_{3}^{-}) = \left( \begin{array}{ccc|ccc|ccc}
 & & \omega& & & & & & \\
 \omega& & & & & & & & \\
 & \omega& & & & & & & \\
 \hline
& & & & \omega^{*}& & & & \\
& & & & &\omega^{*} & & & \\
& & &\omega^{*} & & & & &\\
\hline
& & & & & & & \omega& \\
& & & & & & & &\omega \\
& & & & & &\omega & & \\
\end{array} \right)
\end{equation}

\begin{equation}
    T_{K}(\sigma_{v1}) = \left( \begin{array}{ccc|ccc|ccc}
 & \omega^{*}& & & & & & & \\
 \omega& & & & & & & & \\
 & &1 & & & & & & \\
 \hline
& & & & & & & -1& \\
& & & & & & -1& & \\
& & & & & & & &-1\\
\hline
& & & & -1& & & & \\
& & & -1& & & & & \\
& & & & & -1& & & \\
\end{array} \right)
\end{equation}

\begin{equation}
    T_{K}(\sigma_{v2}) = \left( \begin{array}{ccc|ccc|ccc}
 1& & & & & & & & \\
 & & \omega^{*}& & & & & & \\
 &\omega & & & & & & & \\
 \hline
& & & & & & & &-\omega \\
& & & & & & & -\omega& \\
& & & & & & -\omega& &\\
\hline
& & & & & -\omega^{*}& & & \\
& & & & -\omega^{*}& & & & \\
& & & -\omega^{*}& & & & & \\
\end{array} \right)
\end{equation}

\begin{equation}
    T_{K}(\sigma_{v3}) = \left( \begin{array}{ccc|ccc|ccc}
 & & \omega& & & & & & \\
 & 1& & & & & & & \\
 \omega^{*}& & & & & & & & \\
 \hline
& & & & & &-\omega^{*} & & \\
& & & & & & & & -\omega^{*}\\
& & & & & & & -\omega^{*}&\\
\hline
& & & -\omega& & & & & \\
& & & & & -\omega& & & \\
& & & &-\omega & & & & \\
\end{array} \right)
\end{equation}

\begin{equation}
    T_{M}(E) = \left( \begin{array}{ccc|ccc|ccc}
 1& & & & & & & & \\
 & 1& & & & & & & \\
 & & 1& & & & & & \\
 \hline
& & & 1& & & & & \\
& & & & 1& & & & \\
& & & & & 1& & &\\
\hline
& & & & & & 1& & \\
& & & & & & & 1& \\
& & & & & & & & 1\\
\end{array} \right)
\end{equation}

\begin{equation}
    T_{M}(C_{23}') = \left( \begin{array}{ccc|ccc|ccc}
 1& & & & & & & & \\
 & & 1& & & & & & \\
 & 1& & & & & & & \\
 \hline
& & & & & -1& & & \\
& & & & -1& & & & \\
& & & -1& & & & &\\
\hline
& & & & & & & &-1 \\
& & & & & & & -1& \\
& & & & & &-1 & & \\
\end{array} \right)
\end{equation}

\begin{equation}
    T_{M}(I) = \left( \begin{array}{ccc|ccc|ccc}
1 & & & & & & & & \\
 & -1& & & & & & & \\
 & & -1& & & & & & \\
 \hline
& & & & & & -1& & \\
& & & & & & & -1& \\
& & & & & & & &-1\\
\hline
& & & -1& & & & & \\
& & & & -1& & & & \\
& & & & & -1& & & \\
\end{array} \right)
\end{equation}

\begin{equation}
    T_{M}(\sigma_{v2}) = \left( \begin{array}{ccc|ccc|ccc}
1 & & & & & & & & \\
 & & -1& & & & & & \\
 & -1& & & & & & & \\
 \hline
& & & & & & & & 1\\
& & & & & & & 1& \\
& & & & & & 1& &\\
\hline
& & & & &1 & & & \\
& & & & 1& & & & \\
& & & 1& & & & & \\
\end{array} \right)
\end{equation}

\bibliography{main}

\end{document}